\documentclass[a4paper,english]{article}

\usepackage[margin=2.5cm]{geometry}

\usepackage[table]{xcolor}

\usepackage[hidelinks]{hyperref}

\usepackage{amsmath,mathtools,amssymb,amsfonts,graphicx,caption,subcaption,multicol,multicol,array}

\usepackage[round]{natbib}

\DeclareMathOperator*{\argmax}{arg\,max}

\let\originalleft\left
\let\originalright\right
\renewcommand{\left}{\mathopen{}\mathclose\bgroup\originalleft}
\renewcommand{\right}{\aftergroup\egroup\originalright}

\title{On the Complexity and Behaviour of Cryptocurrencies\\Compared to Other Markets}

\author{Daniel Wilson-Nunn\textsuperscript{1} and Hector Zenil\textsuperscript{2,3,}\footnote{Corresponding author: \href{mailto:hector.zenil@algorithmicnaturelab.org}{\nolinkurl{hector.zenil@algorithmicnaturelab.org}} Both authors contributed equally to this work.}\\
\textsuperscript{1} Department of Statistics, University of Warwick, United Kingdom\\
\textsuperscript{2} Department of Computer Science, University of Oxford, United Kingdom\\
\textsuperscript{3} Algorithmic Nature Group, LABoRES, Paris, France.}

\begin{document}

\nocite{*}

\maketitle

\begin{abstract}
We show that the behaviour of Bitcoin has interesting similarities to stock and precious metal markets, such as gold and silver. We report that whilst Litecoin, the second largest cryptocurrency, closely follows Bitcoin's behaviour, it does not show all the reported properties of Bitcoin. Agreements between apparently disparate complexity measures have been found, and it is shown that statistical, information-theoretic, algorithmic and fractal measures have different but interesting capabilities of clustering families of markets by type. The report is particularly interesting because of the range and novel use of some measures of complexity to characterize price behaviour, because of the IRS designation of Bitcoin as an investment property and not a currency, and the announcement of the Canadian government's own electronic currency \emph{MintChip}.\newline

\end{abstract}

%\keywords{Bitcoin; Financial Time Series; Gold; Silver; Cryptocurrency}

%Supply .eps versions of compression and entropy figures - best done by saving the
   % Mathematica output as a .eps file (Right click on the bracket and press save as),
    %then open the .eps file with a text editor and change %%BoundingBox: 0 0 582 247
    %and %%HiResBoundingBox: 0 0 582 247 to %%BoundingBox: 25 0 582 247 and
    %%HiResBoundingBox: 25 0 582 247
    
%==========================================================================================
%==========================================================================================
%
% .eps versions of entropy and compression figures needed
%
%==========================================================================================
%==========================================================================================

\listoffigures

\listoftables

\section{Introduction}

In March 2014, the Internal Revenue Service (IRS)~\citep{irs} in the U.S. declared that for taxation purposes Bitcoin should be considered property and not currency. The need for the IRS to clarify its position on cryptocurrencies such as Bitcoin and Litecoin arose due to their increasing popularity, and the need for more documentation regarding the taxation of payments and transactions involving cryptocurrencies. Bitcoin has risen in popularity as a method of payment by a growing number of companies, from large online corporations to coffee shops, as well as an investment product due to its increasing value and transient astronomical returns (approximately 8400\% in the eleven months from February 2013 to January 2014). Here, we apply different methods of analysis to the historical prices of Bitcoin and Litecoin and compare the results to those obtained when the same analyses are carried out on other investment products and markets. 

We examine the historical price movements of Bitcoin to determine how the Bitcoin price history compares to other financial markets such as precious metals, stock indices and foreign exchange. We determine whether there is any correlation between the movements of the price of Bitcoin and other financial markets. We analyse the departures of the daily returns of Bitcoin from lognormal compared to that of other financial markets. And, using techniques from information theory such as Shannon's entropy~\citep{shannon}, compressibility and algorithmic probability, we determine to which financial market the Bitcoin price history has the greatest similarity. Lastly, we measure the fractal dimension of the price history of each financial market to find (dis)similarities with each other.

% We found that the behaviour of Bitcoin has stronger similarities to property markets, in particular precious metal markets such as gold and silver than to currency markets. We report that whilst Litecoin closely follows Bitcoin's behaviour, Litecoin, as the second largest crypto currency, does not show all the reported properties of Bitcoin.

\section{Methodology}

It has previously been suggested that if the financial markets behave deterministically rather than scholastically, this should lead to algorithmic signals that may be emulated by computational means~\citep{wolfram} and measured with complexity measures~\citep{algozenil,linma,zenil}. Here we implemented a number of statistic and algorithmic tests to compare prices of a selection of markets, particularly currency, stock and precious metal.

\subsection{Scaling \& Filtering Algorithm}

\label{sec:correlation_algorithm}

The following algorithm is used to scale different time series to the same time period and filter to obtain equal length time series containing the same number of data points.

\begin{description}

\item[Step 1] Pick first point in the cryptocurrency's price history equipped with the converted timestamps and truncated to overlap with the timescale of either gold or silver

\item[Step 2] Set $\mathtt{cointime(1)}$ to equal the timestamp from the first point, and set
$$
\mathtt{metaltime(1)}=\argmax_{\mathtt{x}\in T}\left|\mathtt{x-cointime(1)}\right|
$$
%$$
%\mathtt{metaltime(1)}=\mathrm{argmax}_{\mathtt{x}\in T}\left|\mathtt{x-bitcointime(1)}\right|
%$$
where $T=\left\{\text{All timestamps of the metal's price history}\right\}$.

\item[Step 3] Pick the next (\verb¬i¬-th) point in the cryptocurrency price history equipped with the converted timestamps and truncated to overlap with the timescale of either gold or silver

\item[Step 4] Set $\mathtt{cointime(i)}$ to equal the timestamp from the \verb¬i¬-th point, and set
$$
\mathtt{metaltime(i)}=\argmax_{\mathtt{x}\in T}\left|\mathtt{x-cointime(i)}\right|
$$
%$$
%\mathtt{metaltime(i)}=\mathrm{argmax}_{\mathtt{x}\in T}\left|\mathtt{x-bitcointime(i)}\right|
%$$
where $T$ is as before.

\item[Step 5] Repeat \textbf{Steps 3} and \textbf{4} until all the cryptocurrency's points have been used.

\item[Step 6] Return a list of all the datapoints from the metal that have had their timestamp output.

\end{description}

Using the Bitcoin price history and the filtered precious metal price history, find the correlation using the \verb¬Correlation¬ function in \textit{Mathematica}.

\subsection{Daily Returns}

A common tool for comparing investment products is the analysis of the daily returns of the product. The daily return for day $\mathtt{n}$ is defined as
\begin{equation}
\mathtt{return\left(n\right)}=\frac{\mathtt{price\left(n\right)}}{\mathtt{price\left(n-1\right)}}.	\label{eqn:returns}
\end{equation}

Stock prices have been traditionally modelled using a lognormal distribution. As such, we provide an analysis of the returns of Bitcoin and Litecoin compared to the departure from expected lognormal returns and daily returns of other markets.

\subsection{Entropy and Compressibility}

Shannon's entropy can be written as
\begin{equation}
H\left(X\right)=-\sum_{i=1}^n P\left(x_i\right) \log_2 P\left(x_i\right),
\end{equation}
where $X$ is a random variable with $n$ possible outcomes $\left\{x_1,\dotsc,x_n\right\}$ and $P\left(x_i\right)$ the probability of $x_i$ to occur by an underlying process. In this experiment, price movements are encoded by a binary digit encoding an increase or decrease in price hence the random variable can only take $n=2$ values.

The information content of a message encoded by a single symbol is 0, or in other words, no uncertainty exists in the process of revealing the same outcome bit after bit, because the certainty of the next bit to be zero given its previous history is 1. Hence the entropy of the price movement time series is how uncertain is the next price movement given its history.

Based upon algorithmic results suggested in \citep{zenil} and \citep{linma} we have also applied (algorithmic) information-theoretic measures to price movement data in order to determine their (dis)similarities. A more powerful measure of information and randomness than Shannon's information entropy is provided by the so-called Kolmogorov complexity, $K$. This is because $K$ has been proven to be a universal measure theoretically guaranteed to asymptotically find any computable regularity~\citep{solomonoff1,solomonoff2} in a dataset. Formally, the Kolmogorov complexity~\citep{kolmo,chaitin,solomonoff1,solomonoff2} (also known as Program-size or Kolmogorov-Chaitin complexity) of a string $s$ is given by
\begin{equation}
K\left(s\right)=\min\left\{\left|p\right| : U(p)=s\right\}.
\end{equation}

That is, the length (in bits) of the shortest program $p$ running on a (prefix-free) universal Turing machine, $U$, that outputs $s$ upon halting. By the \emph{Invariance Theorem}~\citep{li}, $K_U$ only depends on $U$ up to a constant, so as is conventional, the $U$ subscript may be dropped.

Despite the inconvenience of its upper semi-computability (only upper bounds can be calculated), $K$ can be effectively approximated by using, for example, lossless compression algorithms, that is compression algorithms for which decompression fully recovers the original object with no loss of information.  In this project we used the compress algorithm as based on Lempel--Ziv--Welch (LZW) compression, a \emph{universal} lossless data compression algorithm.

\subsection{Algorithmic Probability}

The concept of algorithmic probability~\citep{solomonoff1,solomonoff2,levin,chaitin} (also known as Levin's semi-measure) yields a method for approximating the Kolmogorov complexity of a string by means of an optimal but uncomputable solution to the general problem of epistemological induction as stated by Solomonoff~\citep{solomonoff1,solomonoff2}. For example the probability of occurring bits in a finite discretized sequence of values, such as a time series. The algorithmic probability of a time series $x$ is the probability that a random computer program $p$ will produce $x$ when run on a 1-dimensional tape universal (prefix-free~\citep{calude}) Turing machine, $U$. The probability semi-measure $m\left(x\right)$ is related to Kolmogorov complexity $K\left(x\right)$ in that $m\left(G\right)$ is at least the maximum term in the summation of programs ($m\left(G\right)\geq2^{-K\left(G\right)}$), given that the shortest program carries the greatest weight in the sum. The algorithmic \emph{Coding Theorem}~\citep{cover} further establishes the connection between $m\left(x\right)$ and $K\left(x\right)$ as~\citep{levin}: $\left|-\log_2 m\left(x\right) - K\left(x\right)\right| < c$, where $c$ is a constant independent of $s$. The theorem implies that~\citep{cover,calude} one can estimate the Kolmogorov complexity of a graph from the frequency of production from running random programs by simply rewriting the formula as: $K\left(x^\prime\right)=-\log_2 m\left(x\right) + O\left(1\right)$.

In~\citep{d4}, a technique was advanced for approximating $m\left(x\right)$ (hence $K\left(x\right)$) by means of a function that considers all Turing machines of increasing size (by number of states). Indeed, for small values of $n$ states and $k$ colours (usually 2 colours only), $\mathbb{D}\left(n, k\right)$ is computable for values of the Busy Beaver problem~\citep{rado} that are known, providing a means to numerically approximate the Kolmogorov complexity of small graphs, such as network motifs. The \emph{Coding Theorem} then establishes that graphs produced with lower frequency by random computer programs have higher Kolmogorov complexity, and vice-versa. The method is called the \emph{Block Decomposition Method} (BDM) because it consists of decomposing the time series in smaller (potentially overlapping) subtime series of lengths for which complexity values have been estimated, then reconstructing an approximation of the Kolmogorov complexity of the time series by adding the complexity of local (but overlapping) regularities in the individual subtime series. More formally,
$$
K\left(x\right) = \sum_{{\left(r_u,n_u\right)\in s_{d d}}} \log_2\left(n_u\right)+K\left(r_u\right),
$$
where $s_{d}$ represents the set with elements $\left(r_u,n_u\right)$, obtained when decomposing the time series into subtime series of length $d$. In each $\left(r_u,n_u\right)$ pair, $r_u$ is one such subtime series  and $n_u$ is its multiplicity (number of occurrences in the original time series). Applications of the BDM have been explored in~\citep{d4,d5,numerical,kolmo2d,zenilchaos,BRM}, and include applications to graph theory and complex networks~\citep{zenilgraph}.

\subsection{Fractal Roughness of Markets}

The historical prices of a financial asset can be seen as a fractal, and much research has gone into determining various fractal properties of such time series. One interesting way to compare different time series is to compare their fractal roughness, and work has been done on this topic since inspired by Mandelbrot~\citep{mandelbrot}. The Box-Count estimator is a commonly used method for calculating fractal roughness and it motivated by the scaling law~\eqref{eqn:scalinglaw}.
\begin{equation}
D_{\mathrm{BC}}=\lim_{\varepsilon\to0}\frac{\log{N\left(\varepsilon\right)}}{\log{\left(1/\varepsilon\right)}}	\label{eqn:scalinglaw}
\end{equation}

The method used for calculating the fractal dimension used in this paper is the Hall-Wood estimator~\citep{hall,timeseries} which was introduced in 1993 and is a variation of the Box-Count estimator. To use the Hall-Wood estimator, data has to first be of the form
\begin{equation}
\left\{\left(t,X_t\right):t=\frac{i}{n},i=0,1,\dotsc,n\right\}\subset\mathbb{R}^2.
\end{equation}
As the data is assumed to be daily prices, it is possible to simply set $n$ to be the number of datapoints in each dataset and let each point have the timestamp $i/n,\left(i+1\right)/n,\dotsc$.

The Hall-Wood estimator reformulates the definition of the scaling law~\eqref{eqn:scalinglaw} using the function $A\left(\varepsilon\right)$ to denote the total area of boxes at scale $\varepsilon$ that intersect with the linearly interpolated data graph. As there are $N\left(\varepsilon\right)$ of these boxes, it is easy to deduce that $A\left(\varepsilon\right)\propto N\left(\varepsilon\right)\varepsilon^2$.
\begin{equation}
D_{\mathrm{BC}}=2-\lim_{\varepsilon\to0}\frac{\log{A\left(\varepsilon\right)}}{\log{\left(\varepsilon\right)}}.	\label{eqn:scalinglawredef}
\end{equation}
Setting the scale $\varepsilon_l=l/n$, for $l=1,2,\dotsc$ the value of $A\left(l/n\right)$ can be estimated as
\begin{equation}
\widehat{A}\left(l/n\right)=\frac{l}{n}\sum_{i=1}^{\left\lfloor n/l\right\rfloor}\left|X_{il/n}-X_{\left(i-1\right)l/n}\right|,
\end{equation}
where $\left\lfloor n/l\right\rfloor$ is the floor of $n/l$ where $\mathrm{floor}\left(x\right)=\left\lfloor x\right\rfloor:=\max\left\{k\in\mathbb{Z}:k\leq x\right\}$. Based on an ordinary least squares regression fit of $\log{\widehat{A}\left(l/n\right)}$ on $\log{\left(l/n\right)}$ the following is obtained:
\begin{equation}
\widehat{D}_{\mathrm{HW}}=2-\left\{\sum_{l=1}^L\left(s_l-\overline{s}\right)\log{\widehat{A}\left(l/n\right)}\right\}
\left\{\sum_{l=1}^L\left(s_l-\overline{s}\right)^2\right\}^{-1},
\end{equation}
with
\begin{align*}
L	&\geq	2,	&s_l	&=	\log{\left(l/n\right)}	&\overline{s}	&=	\frac{1}{L}\sum_{l=1}^Ls_l.
\end{align*}

Picking $L=2$ to minimise bias when taking the limit as $\varepsilon\to0$ in~\eqref{eqn:scalinglaw}, gives rise to the Hall-Wood estimator
\begin{equation}
\widehat{D}_{\mathrm{HW}}=2-\frac{\log{\widehat{A}\left(2/n\right)}-\log{\widehat{A}\left(1/n\right)}}{\log{2}}.
\end{equation}

Box-counting dimension~\citep{dauphine} has been strongly criticised because of its scale dependency, but here time series had the same scale so the effect of this downside should be minimised.

\section{Results}

\subsection{Daily Returns Behaviour}

\begin{figure}
	\centering
	\begin{subfigure}{0.45\textwidth}
		\centering
		\makebox{\includegraphics[width=\textwidth]{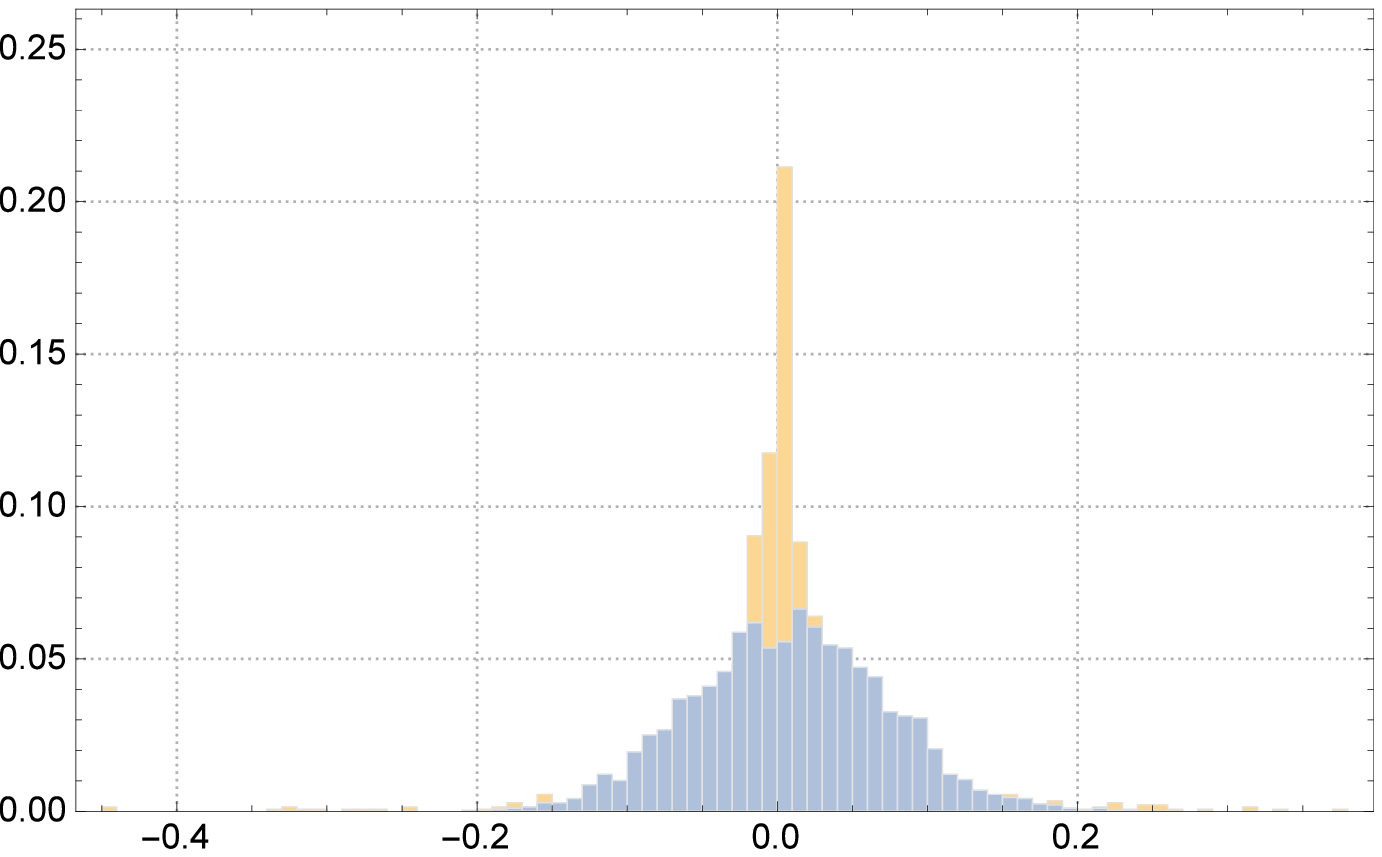}}
		\caption{\label{fig:bitcoin_histogram}Bitcoin returns (orange/light grey) against expected returns (blue/dark grey).}
	\end{subfigure}%
	~
	\begin{subfigure}{0.45\textwidth}
		\centering
		\makebox{\includegraphics[width=\textwidth]{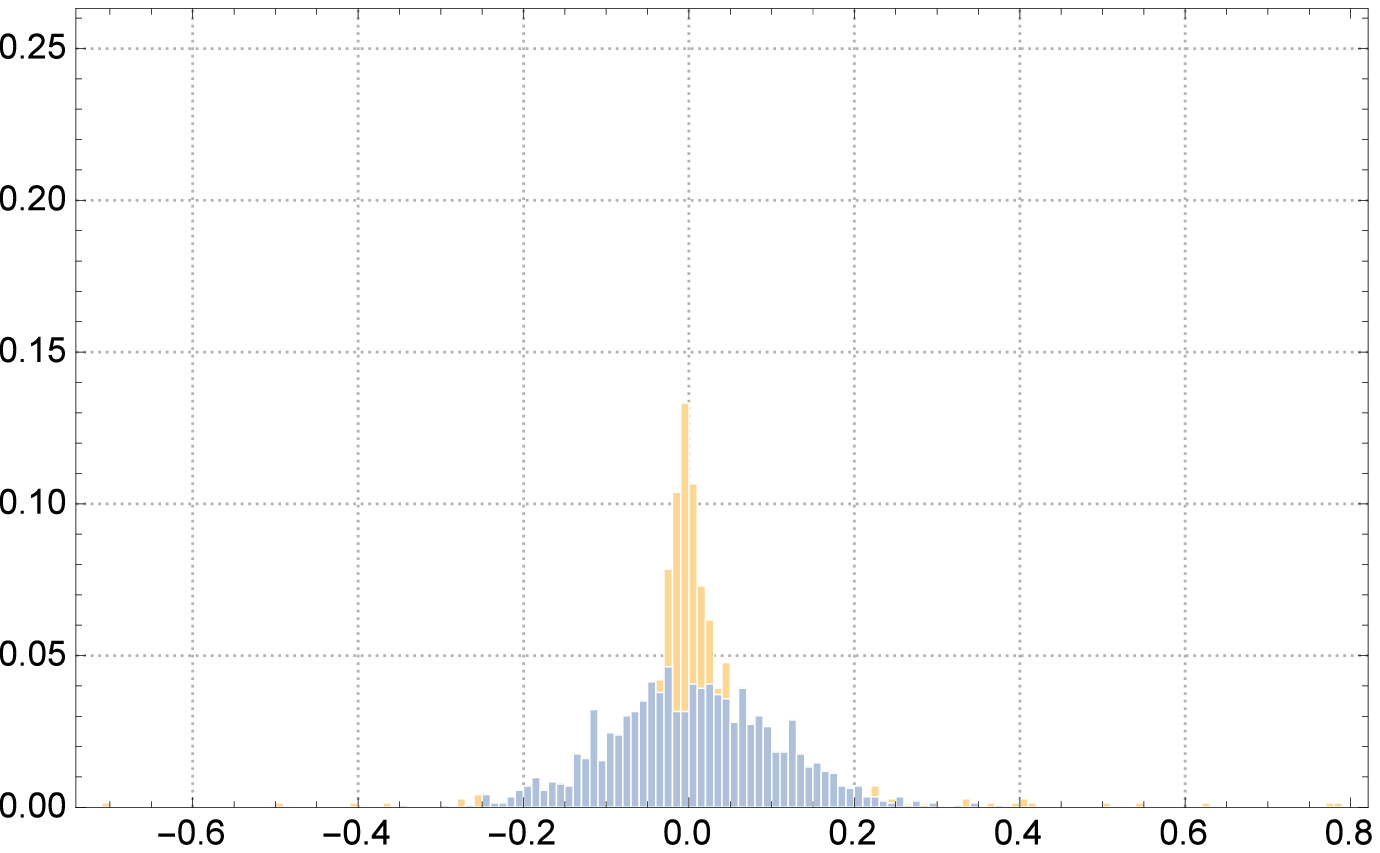}}
		\caption{\label{fig:litecoin_histogram}Litecoin returns (orange/light grey) against expected returns (blue/dark grey).}
	\end{subfigure}
	\caption{\label{fig:crypto_returns}Comparison of logarithm of daily returns of cryptocurrencies to expected returns given a normal distribution.}
\end{figure}

\begin{figure}
	\centering
	\makebox[\linewidth][c]{%
	\begin{subfigure}{0.45\textwidth}
		\centering
		\makebox{\includegraphics[width=\textwidth]{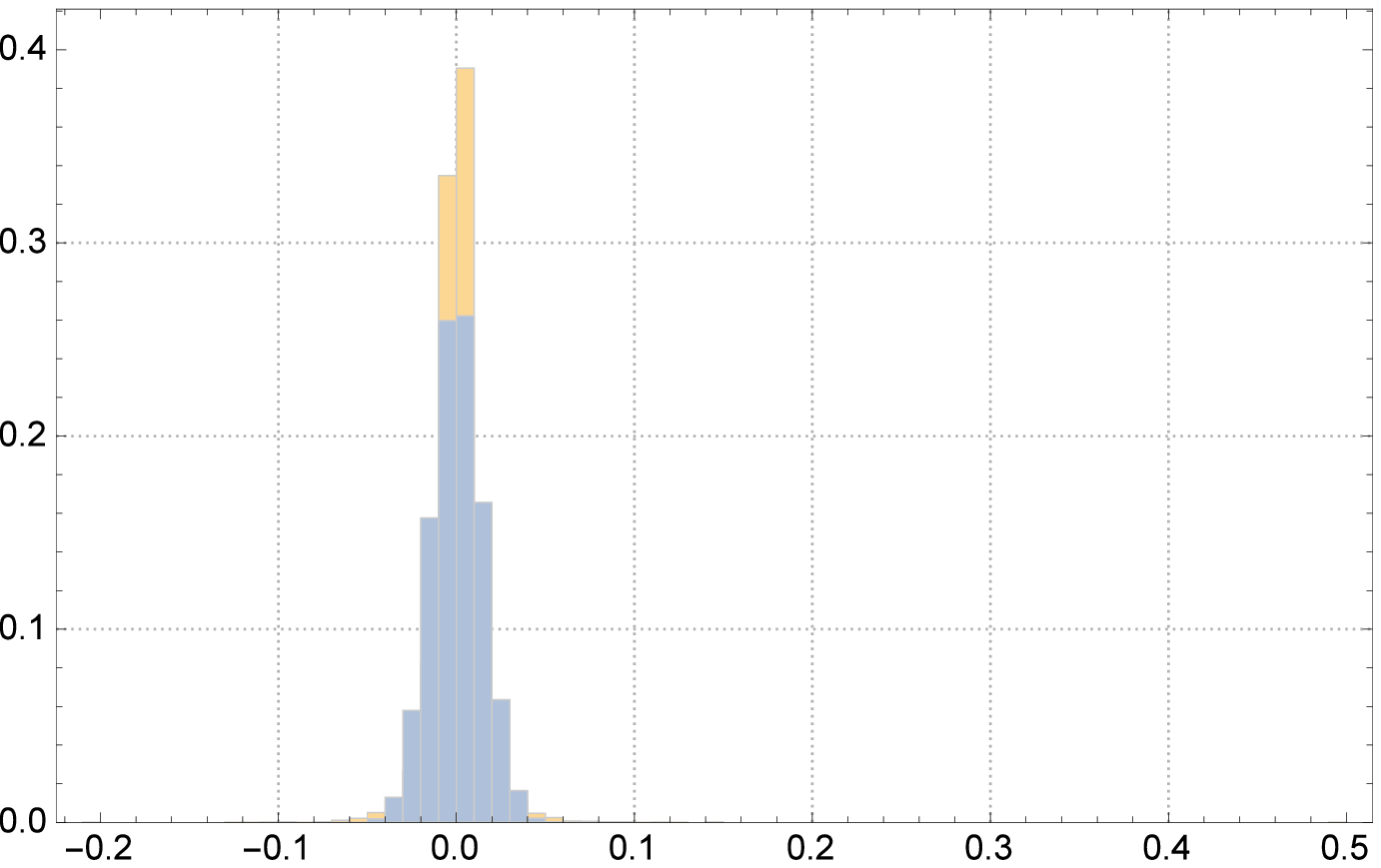}}
		\caption{\label{fig:allgold_histogram}All gold returns (orange/light grey) against expected returns (blue/dark grey).}
	\end{subfigure}%
	~
	\begin{subfigure}{0.45\textwidth}
		\centering
		\makebox{\includegraphics[width=\textwidth]{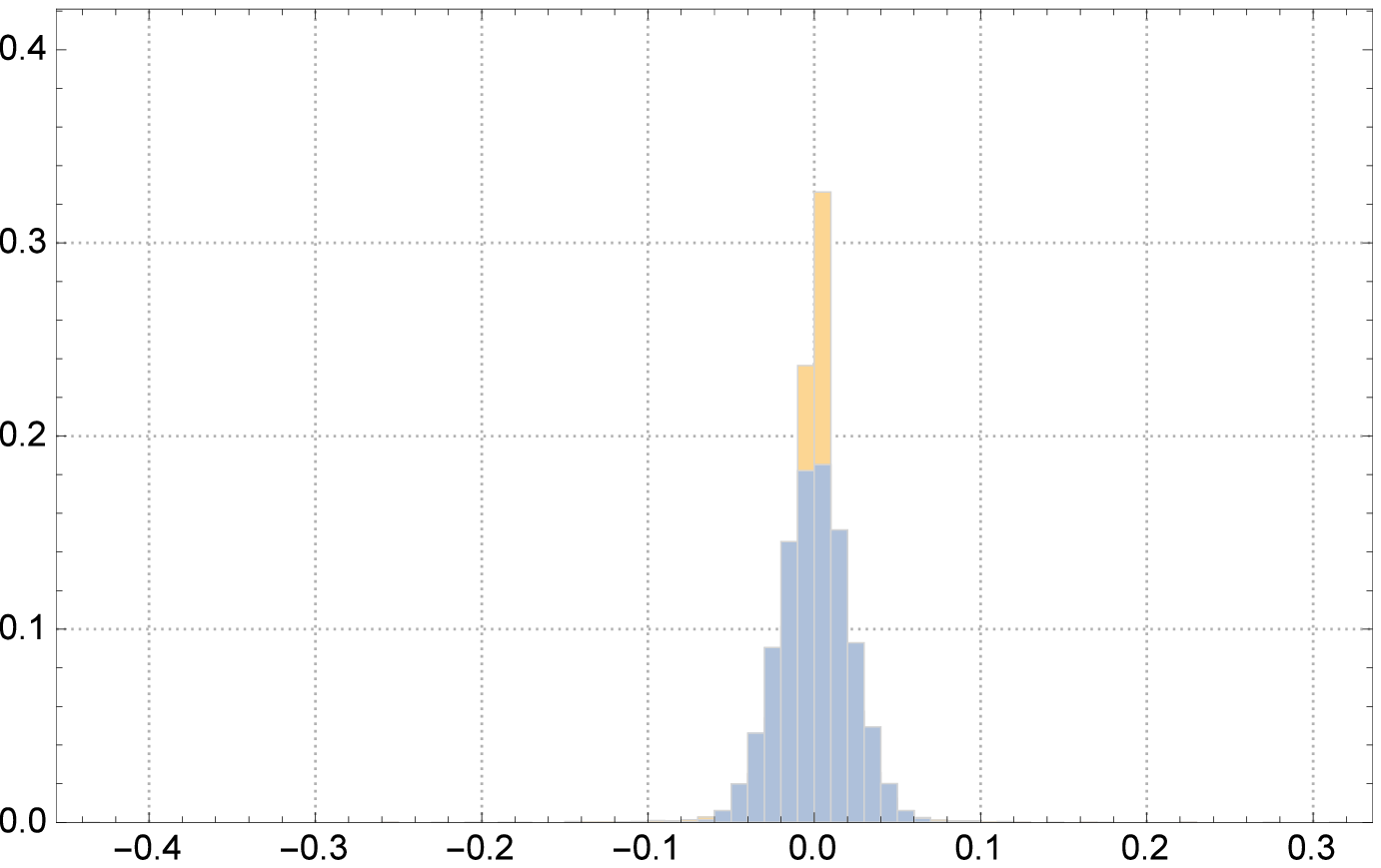}}
		\caption{\label{fig:allsilver_histogram}All silver returns (orange/light grey) against expected returns (blue/dark grey).}
	\end{subfigure}%
	}\\
	\makebox[\linewidth][c]{%
	\begin{subfigure}{0.45\textwidth}
		\centering
		\makebox{\includegraphics[width=\textwidth]{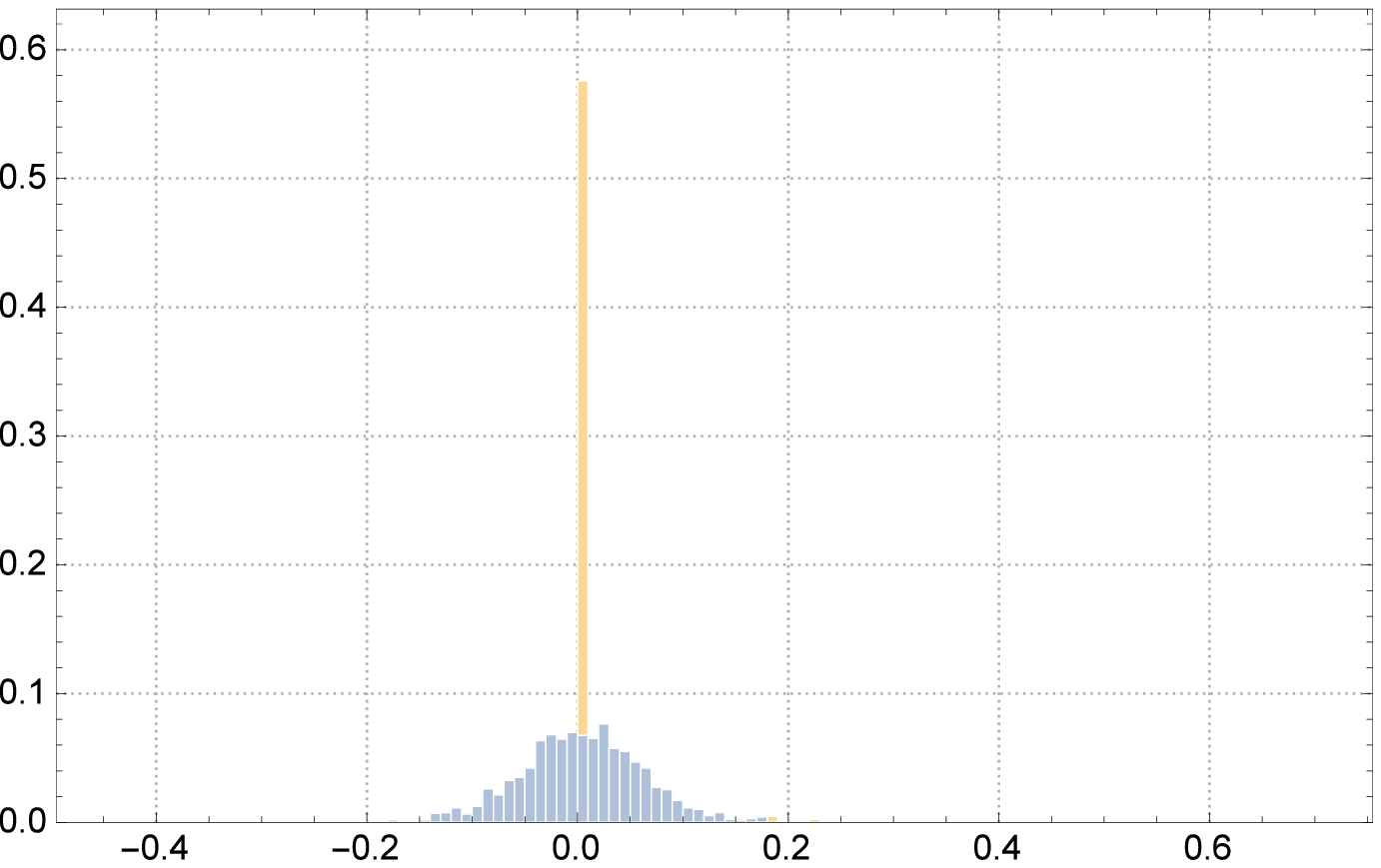}}
		\caption{\label{fig:filteredgold_histogram}Filtered gold returns (orange/light grey) against expected returns (blue/dark grey).}
	\end{subfigure}%
	~
	\begin{subfigure}{0.45\textwidth}
		\centering
		\makebox{\includegraphics[width=\textwidth]{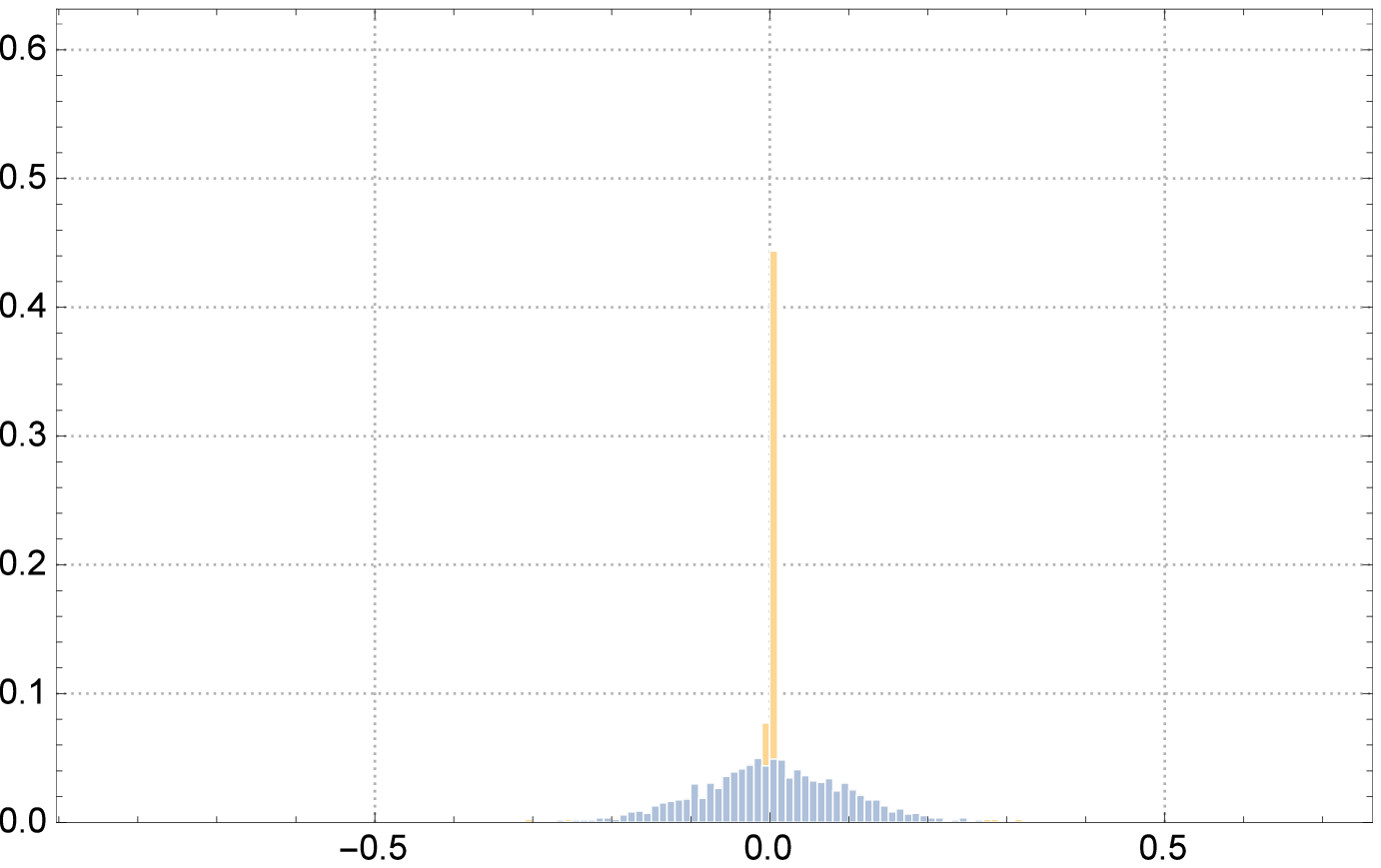}}
		\caption{\label{fig:filteredsilver_histogram}Filtered silver returns (orange/light grey) against expected returns (blue/dark grey).}
	\end{subfigure}
	}\\
	\caption{\label{fig:precious_returns}Comparison of logarithm of daily returns of precious metals to expected returns given a normal distribution.}
\end{figure}

\begin{figure}
	\centering
	\makebox[\linewidth][c]{%
	\begin{subfigure}{0.45\textwidth}
		\centering
		\makebox{\includegraphics[width=\textwidth]{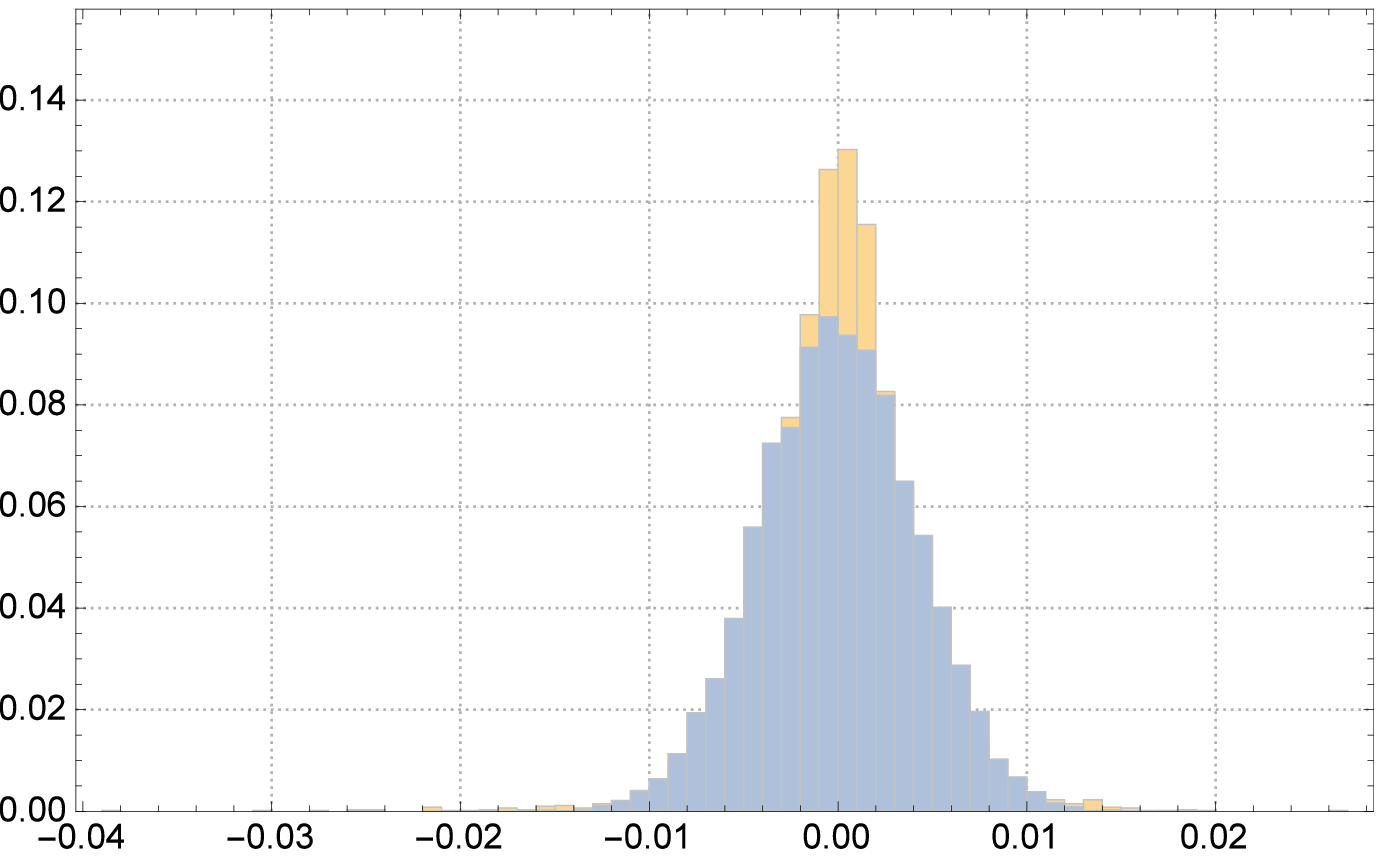}}
		\caption{\label{fig:gbp_histogram}All GBP/USD returns (orange/light grey) against expected returns (blue/dark grey).}
	\end{subfigure}%
	~
	\begin{subfigure}{0.45\textwidth}
		\centering
		\makebox{\includegraphics[width=\textwidth]{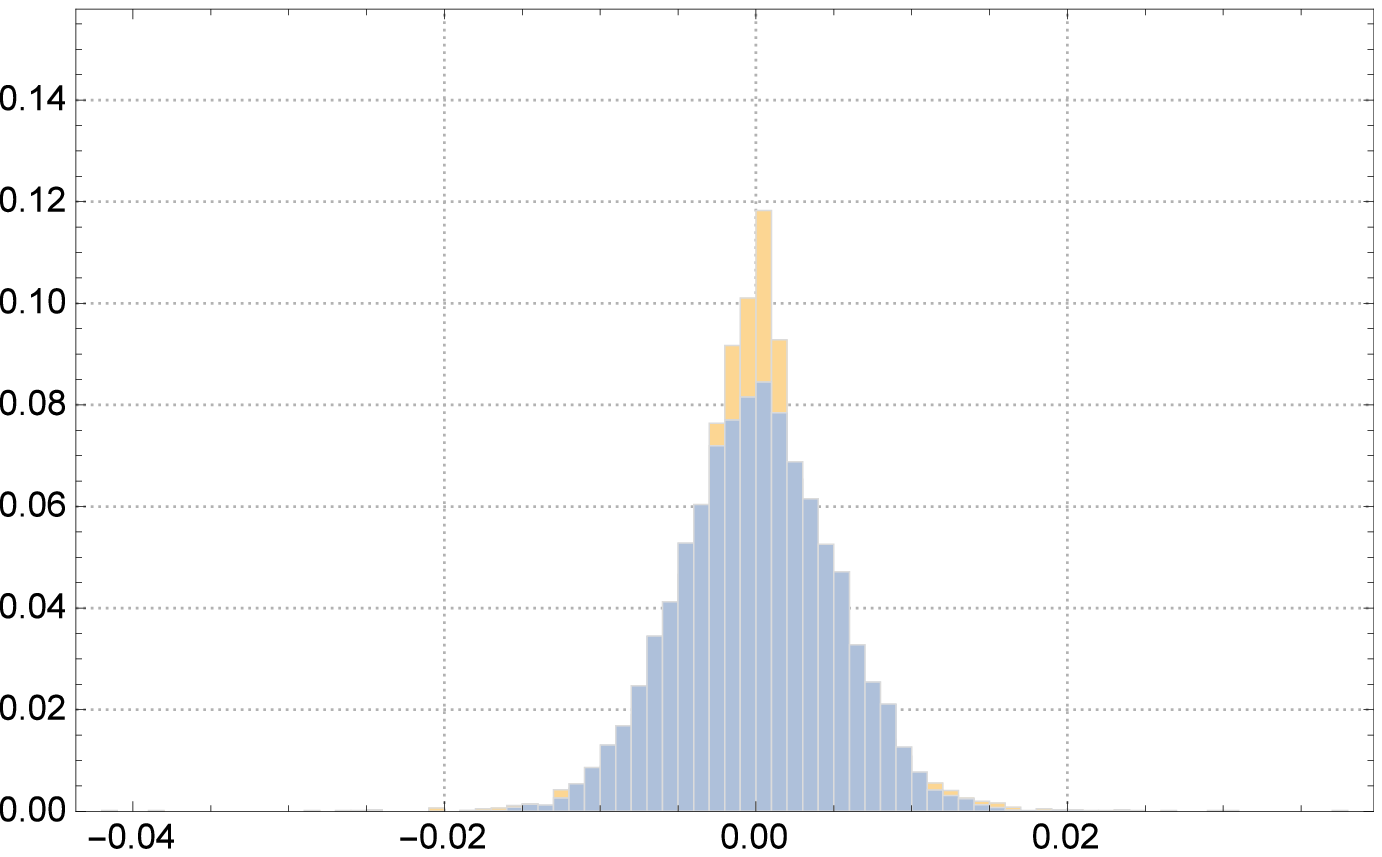}}
		\caption{\label{fig:chf_histogram}All CHF/USD returns (orange/light grey) against expected returns (blue/dark grey).}
	\end{subfigure}%
	}\\
	\makebox[\linewidth][c]{%
	\begin{subfigure}{0.45\textwidth}
		\centering
		\makebox{\includegraphics[width=\textwidth]{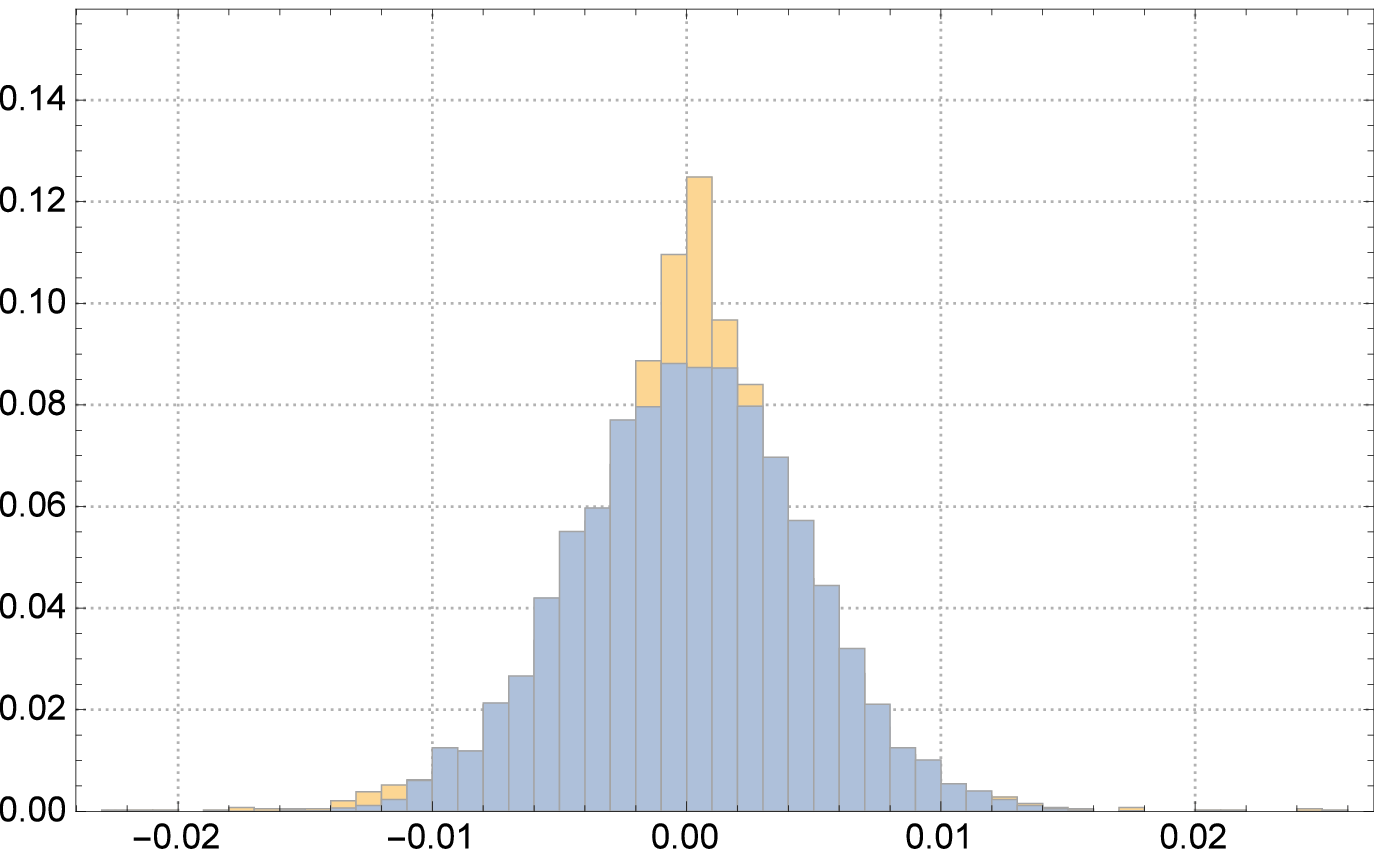}}
		\caption{\label{fig:eur_histogram}All EUR/USD returns (orange/light grey) against expected returns (blue/dark grey).}
	\end{subfigure}
	}\\
	\caption{\label{fig:forex_returns}Comparison of logarithm of daily returns of foreign exchange rates to expected returns given a normal distribution}
\end{figure}

\begin{figure}
	\centering
	\makebox[\linewidth][c]{%
	\begin{subfigure}{0.45\textwidth}
		\centering
		\makebox{\includegraphics[width=\textwidth]{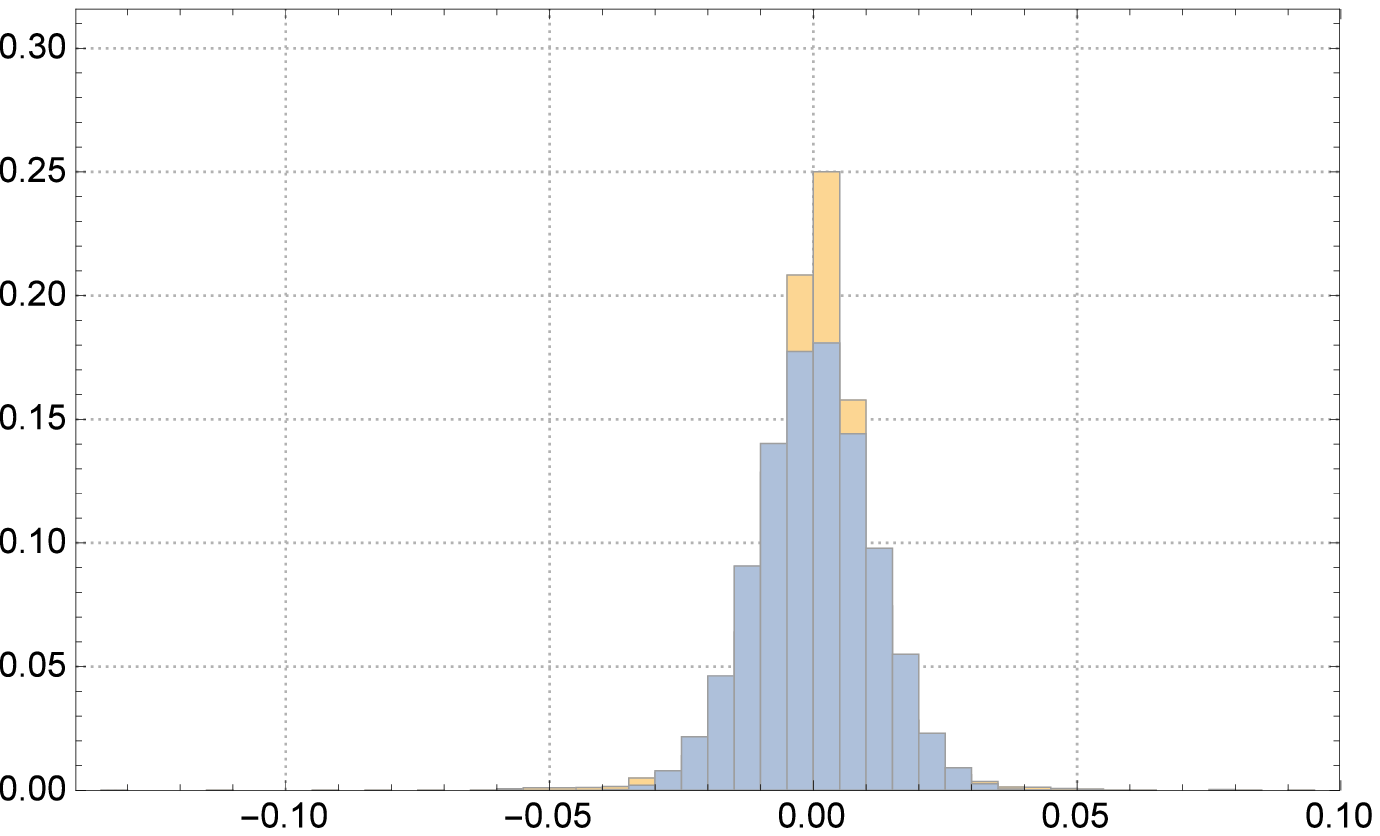}}
		\caption{\label{fig:ftse_histogram}All FTSE100 returns (orange/light grey) against expected returns (blue/dark grey).}
	\end{subfigure}%
	~
	\begin{subfigure}{0.45\textwidth}
		\centering
		\makebox{\includegraphics[width=\textwidth]{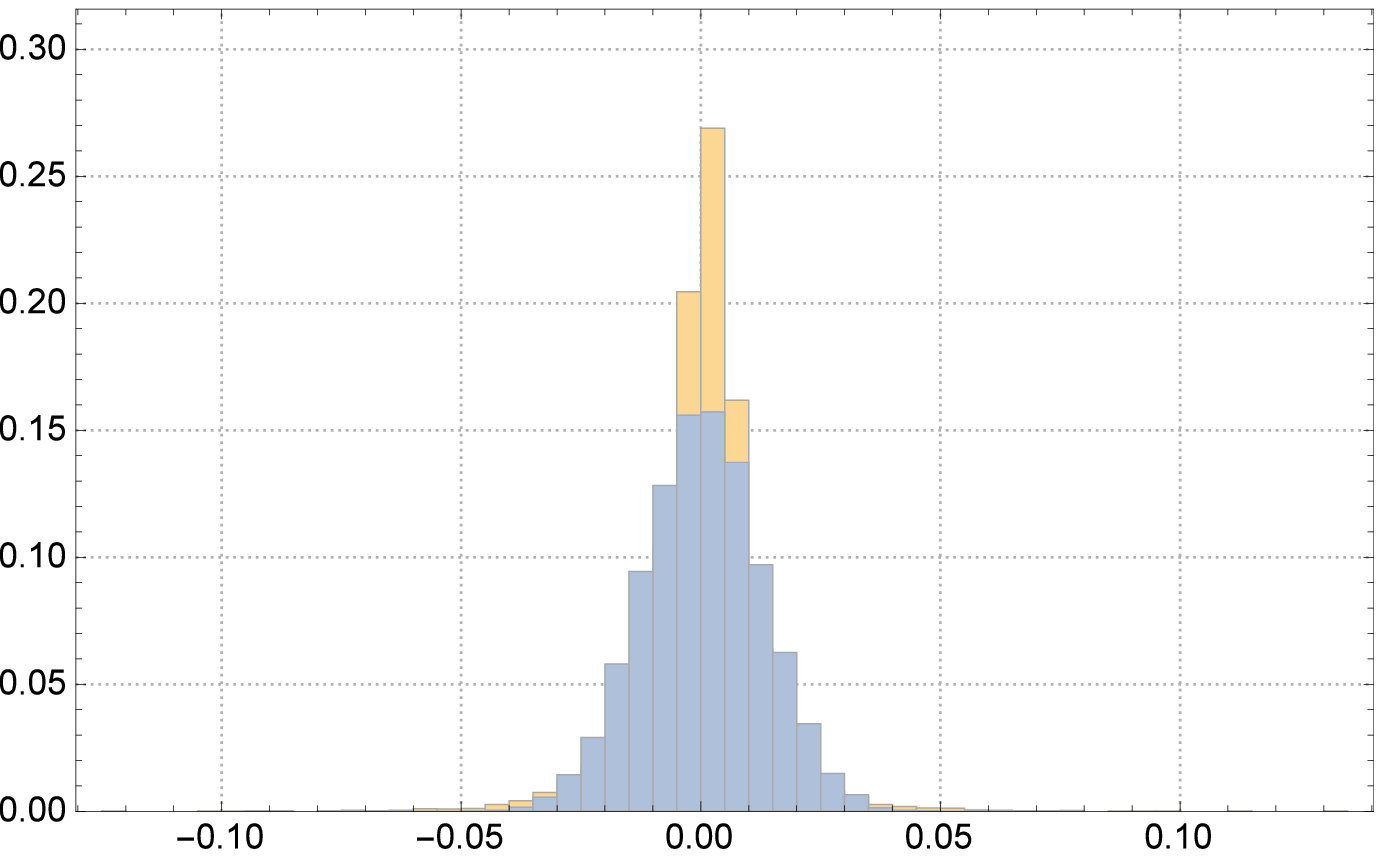}}
		\caption{\label{fig:nasdaq_histogram}All NASDAQ returns (orange/light grey) against expected returns (blue/dark grey).}
	\end{subfigure}%
	}\\
	\makebox[\linewidth][c]{%
	\begin{subfigure}{0.45\textwidth}
		\centering
		\makebox{\includegraphics[width=\textwidth]{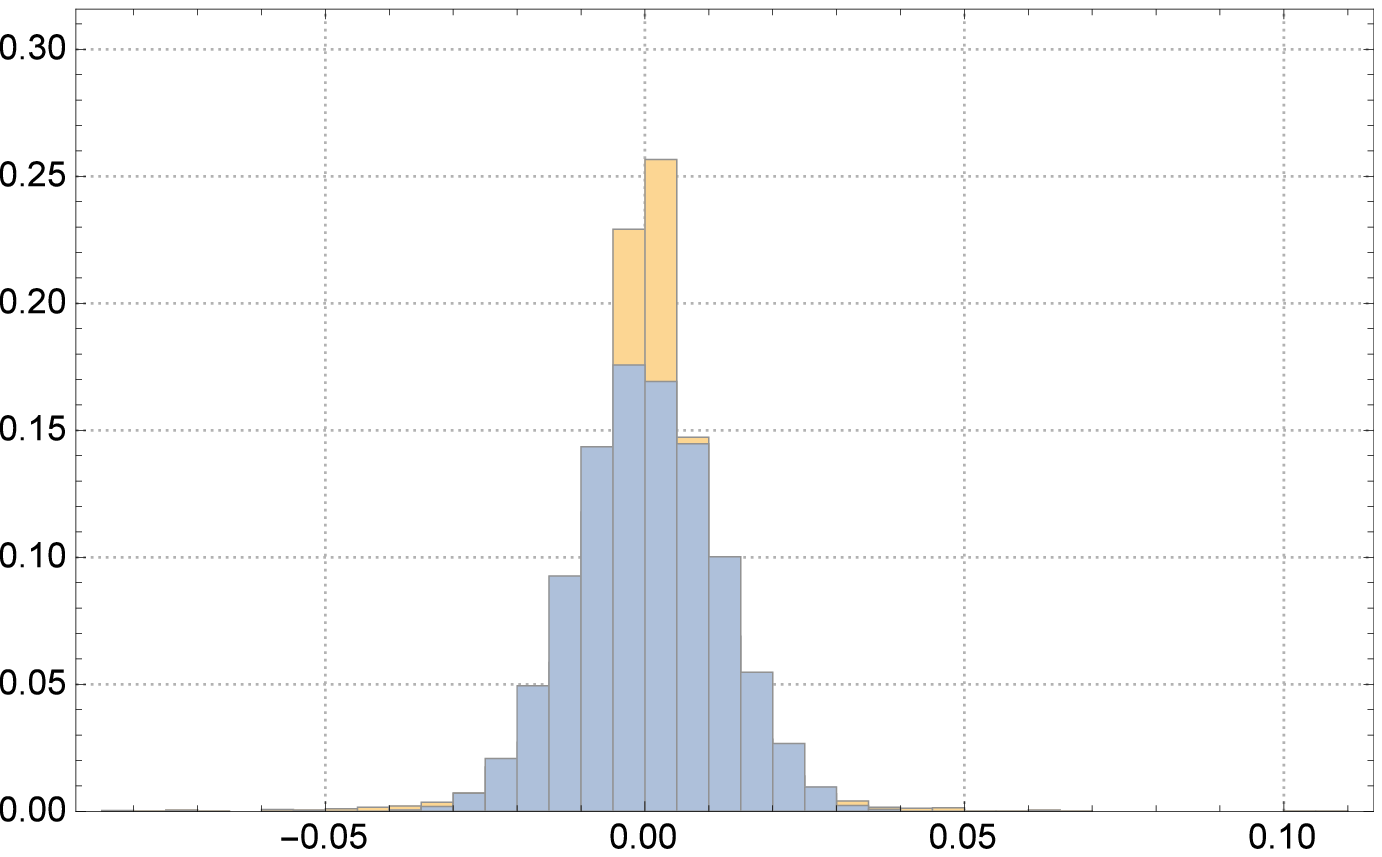}}
		\caption{\label{fig:dow_histogram}All Dow Jones (orange/light grey) returns against expected returns (blue/dark grey).}
	\end{subfigure}%
	~
	\begin{subfigure}{0.45\textwidth}
		\centering
		\makebox{\includegraphics[width=\textwidth]{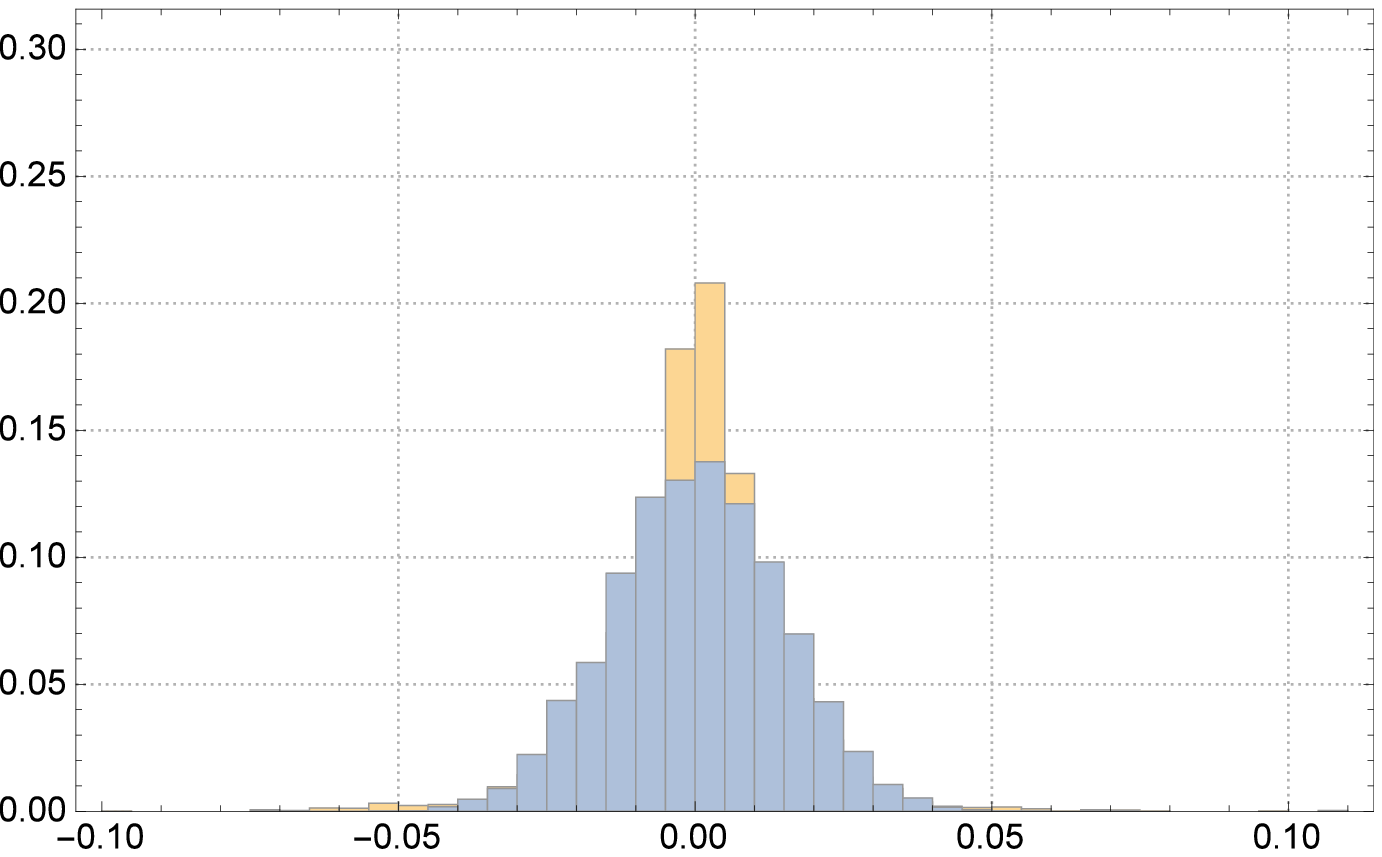}}
		\caption{\label{fig:dax_histogram}All DAX returns (orange/light grey) against expected returns (blue/dark grey).}
	\end{subfigure}%
	}\\
	\makebox[\linewidth][c]{%
	\begin{subfigure}{0.45\textwidth}
		\centering
		\makebox{\includegraphics[width=\textwidth]{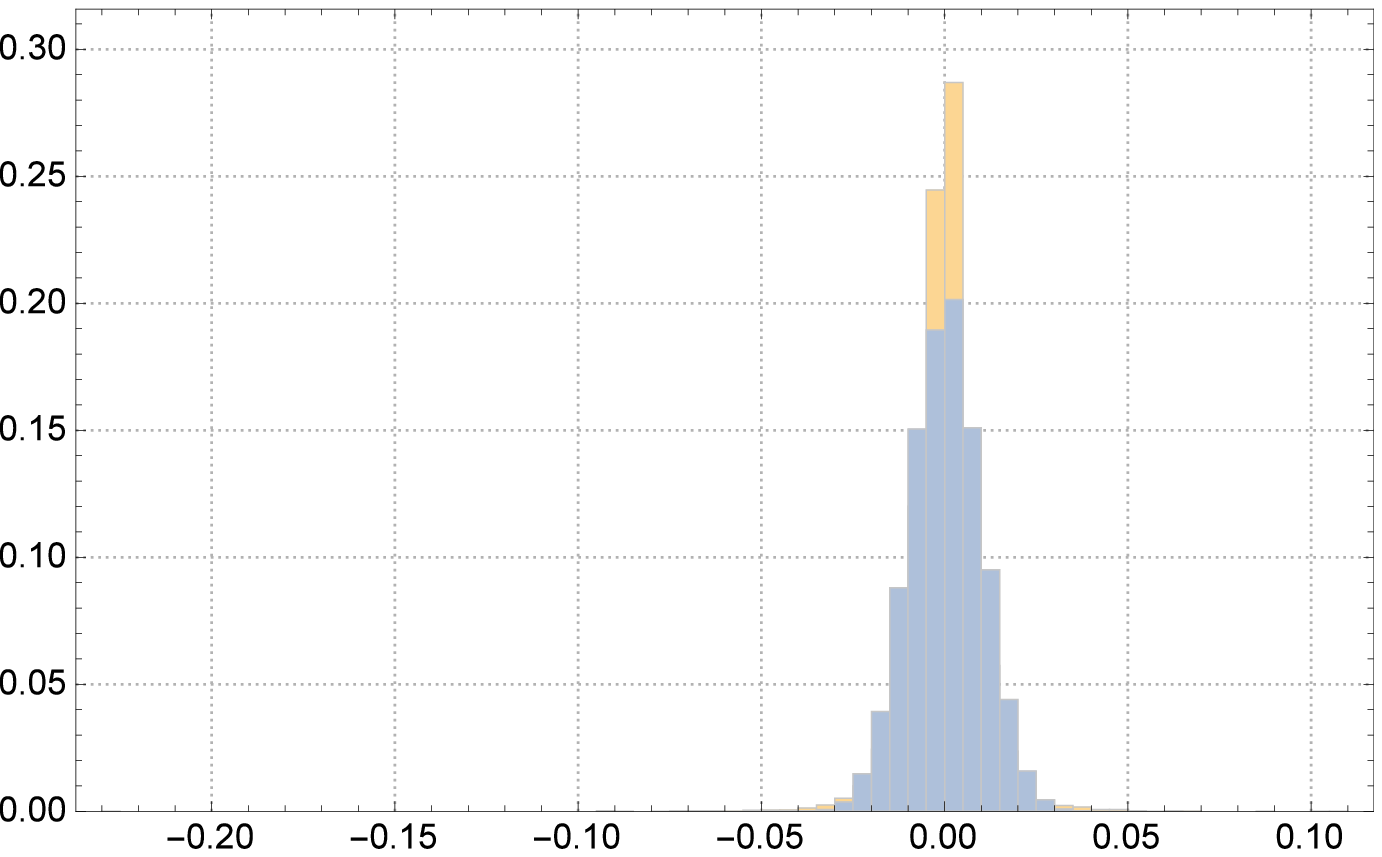}}
		\caption{\label{fig:sandp_histogram}All S\&P500 returns (orange/light grey) against expected returns (blue/dark grey).}
	\end{subfigure}
	}\\
	\caption{\label{fig:index_returns}Comparison of logarithm of daily returns of stock indices to expected returns given a normal distribution.}
\end{figure}

Upon calculating the logarithm of the daily returns for each financial product from the set of possible similar financial products in Table~\ref{tab:comps}, key statistics about the product can be easily compared. Table~\ref{tab:returns} shows the products along with key statistics about the logarithm of their daily returns. Comparing the logarithm of the daily returns of each product to that expected if the returns followed the lognormal distribution can be done visually using histograms as in Figs.~\ref{fig:crypto_returns}~-~\ref{fig:index_returns}.

{\begin{table}
\centering
\fbox{
\begin{tabular}{lllll}
							&	\emph{Mean}	&	\emph{Standard deviation}	&	\emph{Kurtosis}	&	\emph{Skewness}	\\ \hline
	\textbf{Bitcoin}		&	0.00658267		&	0.0655309						&	11.9791				&	-0.300827			\\
	\textbf{Litecoin}		&	0.00760243		&	0.103948						&	19.5837				&	1.46262				\\
	\textbf{Gold}			&	0.000354129	&	0.01406							&	150.538				&	3.77226				\\
	\textbf{Silver}		&	0.000265421	&	0.0209999						&	35.8672				&	-0.738693			\\
	\textbf{GBP/USD}		&	-0.0000171049	&	0.0040977						&	8.20226				&	-0.551858			\\
	\textbf{CHF/USD}		&	0.0000666324	&	0.00489236						&	7.24448				&	0.0123075			\\
	\textbf{EUR/USD}		&	0.0000651475	&	0.00447983						&	5.01895				&	-0.0586363			\\
	\textbf{FTSE100}		&	0.000242546	&	0.0109507						&	12.9839				&	-0.495725			\\
	\textbf{NASDAQ	}		&	0.000347418	&	0.0124925						&	12.8056				&	-0.293667			\\
	\textbf{Dow Jones}	&	0.000296831	&	0.0110213						&	11.5196				&	-0.166408			\\
	\textbf{DAX}			&	0.000324354	&	0.0143508						&	7.79474				&	-0.108969			\\
	\textbf{S\&P500}		&	0.000294571	&	0.00973758						&	30.715				&	-1.03097			\\
\end{tabular}}
\caption{\label{tab:returns}Analysis of the logarithm of the daily returns of markets.}
\end{table}}

{\begin{table}
\centering
\fbox{
\begin{tabular}{lll}
						&	\emph{Start Date}	&	\emph{Product Type}	\\ \hline
\textbf{Bitcoin}		&	17/07/2010			&	Cryptocurrency		\\
\textbf{Litecoin}		&	13/07/2012			&	Cryptocurrency		\\
\textbf{Gold}			&	01/03/1900			&	Precious metal			\\
\textbf{Silver}		&	01/03/1900			&	Precious metal			\\
\textbf{GBP/USD}		&	05/03/1991			&	Foreign exchange		\\
\textbf{CHF/USD}		&	05/03/1991			&	Foreign exchange		\\
\textbf{EUR/USD}		&	06/09/1999			&	Foreign exchange		\\
\textbf{FTSE100}		&	03/01/1984			&	Stock index			\\
\textbf{NASDAQ	}		&	05/02/1971			&	Stock index			\\
\textbf{Dow Jones}	&	02/01/1992			&	Stock index			\\
\textbf{DAX}			&	26/11/1990			&	Stock index			\\
\textbf{S\&P500}		&	03/01/1950			&	Stock index			\\
\end{tabular}}
\caption{\label{tab:comps}Financial products used in this paper as possible comparables to cryptocurrencies.}
\end{table}}

It is clear in both Table~\ref{tab:returns} and Figs.~\ref{fig:crypto_returns}~-~\ref{fig:index_returns} that there is distinct grouping of the financial instruments by the properties of their returns. By comparing both the means and the standard deviations of the logarithm of the daily returns, three distinct classes of financial product are formed; one class containing the cryptocurrencies, another containing the foreign exchanges and a final class containing precious metals and stock indices. In each case, the behaviour of the cryptocurrencies is much closer to precious metals and stock indices than to foreign exchange.

\subsection{Correlation Tests}

\begin{figure}
	\centering
	\begin{subfigure}{0.45\textwidth}
		\centering
		\makebox{\includegraphics[width=\textwidth]{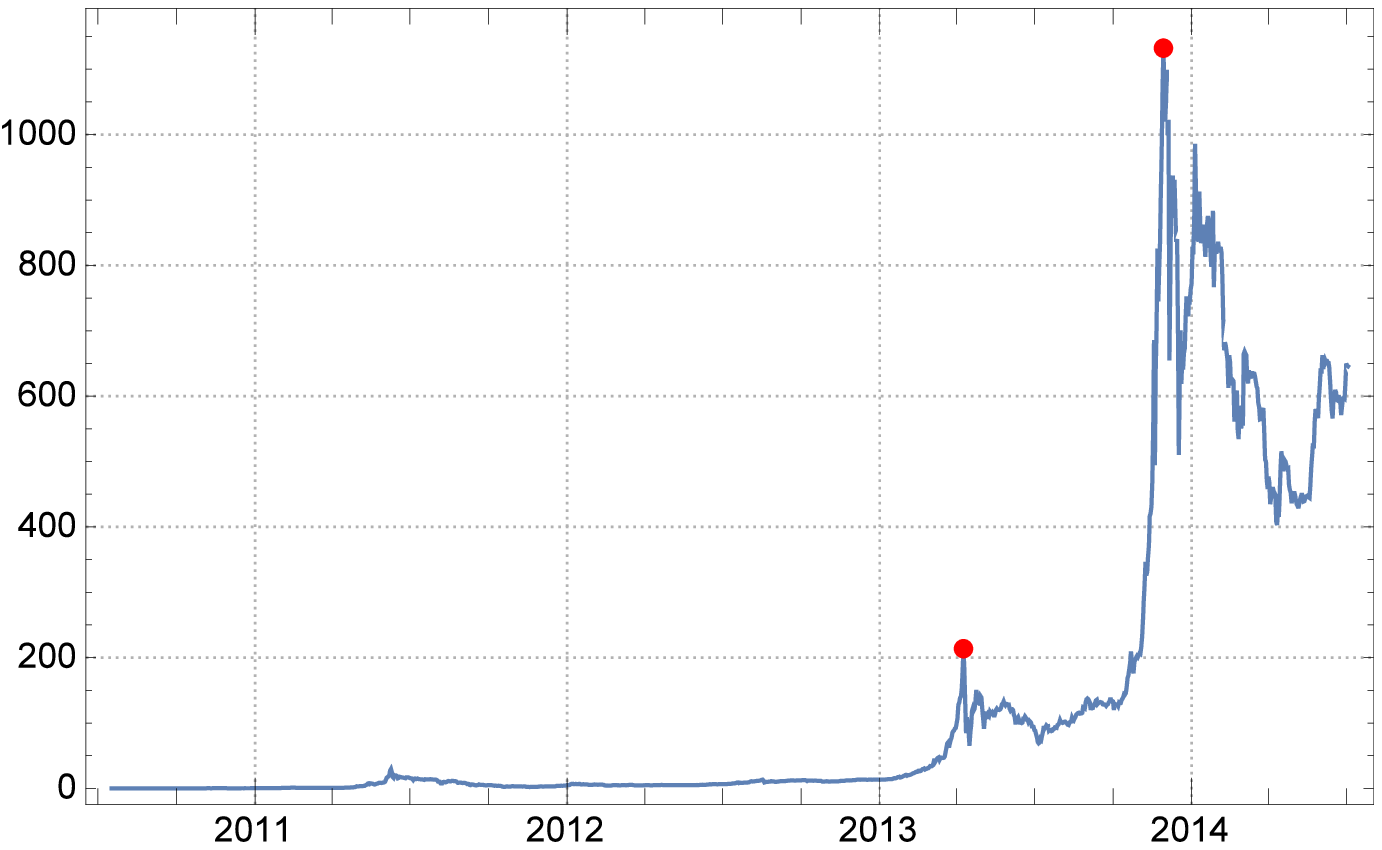}}
		\caption{\label{fig:bitcoin}Historical price of Bitcoin.}
	\end{subfigure}%
	~
	\begin{subfigure}{0.45\textwidth}
		\centering
		\includegraphics[width=\textwidth]{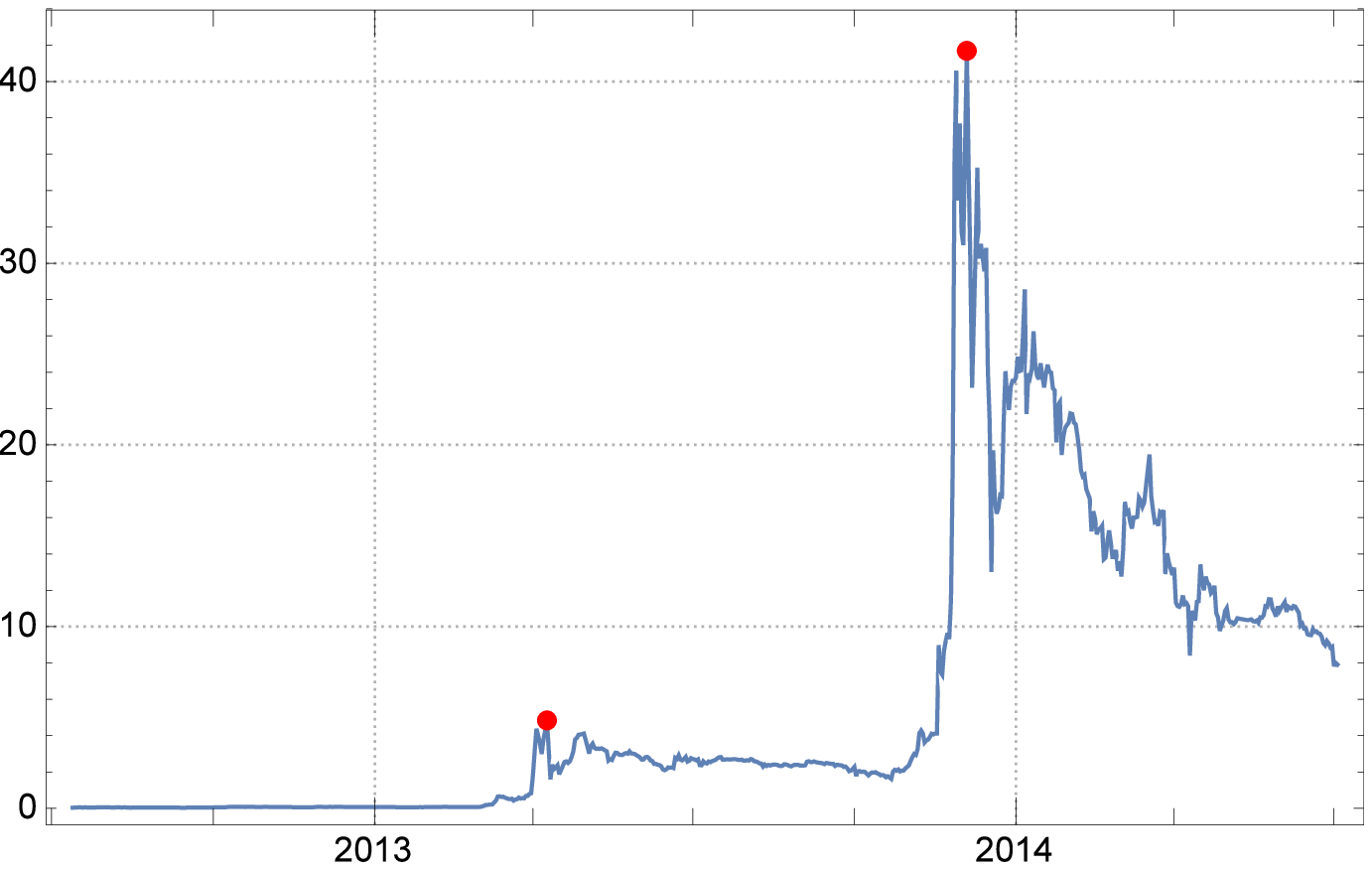}
		\caption{\label{fig:litecoin}Historical price of Litecoin.}
	\end{subfigure}
	\caption{\label{fig:crypto}Historical price of cryptocurrencies with main peaks highlighted.}
\end{figure}

\begin{figure}
	\centering
	\begin{subfigure}{0.45\textwidth}
		\centering
		\makebox{\includegraphics[width=\textwidth]{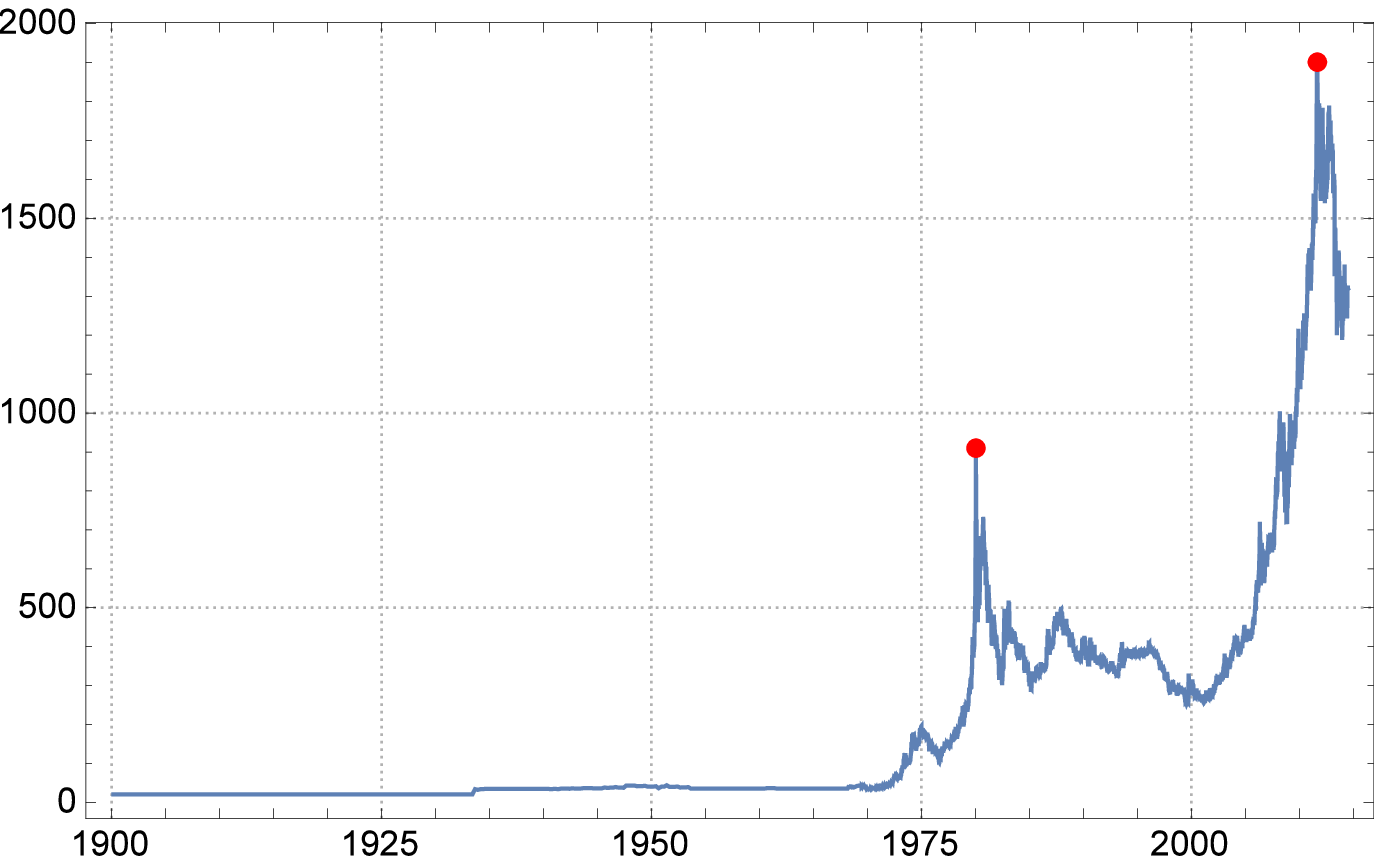}}
		\caption{\label{fig:gold}Historical price of gold.}
	\end{subfigure}%
	~
	\begin{subfigure}{0.45\textwidth}
		\centering
		\includegraphics[width=\textwidth]{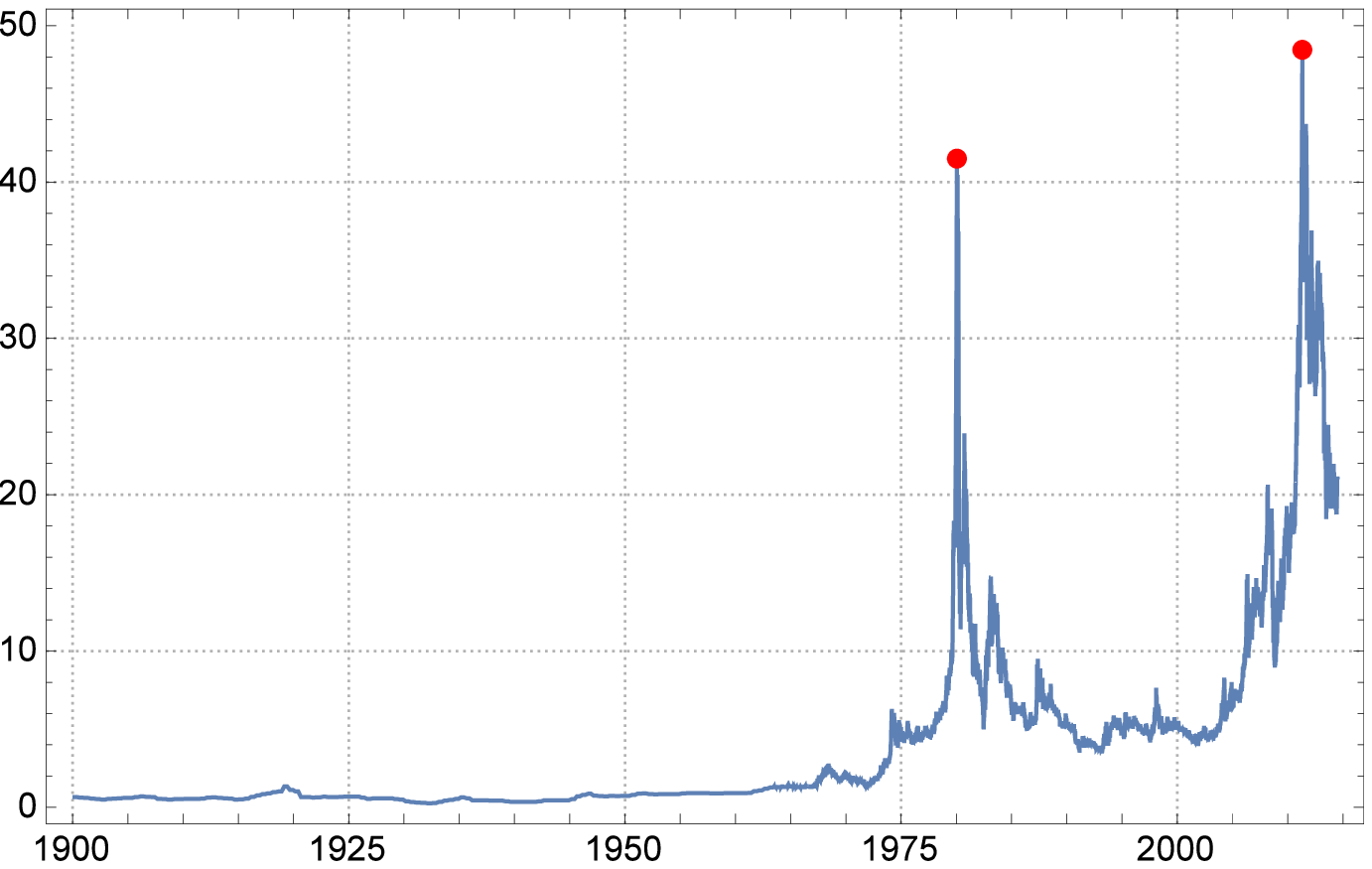}
		\caption{\label{fig:silver}Historical price of silver.}
	\end{subfigure}
	\caption{\label{fig:precious}Historical price of precious metals with main peaks highlighted.}
\end{figure}

Inspecting the historical price of Bitcoin, Litecoin, gold and silver in Figs.~\ref{fig:crypto}~and~\ref{fig:precious}, it appears there is a similar shape that the three investment products follow. To determine whether this visual similarity maps to a statistically significant correlation between the historic prices of Bitcoin, gold and silver, it is necessary to determine the correlation between upwards and downwards movements of the price.

To achieve this it was first necessary to scale the historical prices of Bitcoin, Litecoin, gold and silver to be over the same time period. To do this, dates were converted into absolute time (the time since 01/01/1900 00:00 in seconds, and can be found in \textit{Wolfram Mathematica} using the \texttt{AbsoluteTime} command) and then scaled to ensure that the notable features (peaks) were aligned using a linear transformation. 
%This conversion and transformation leads to the data becoming of the form seen in Figure~\ref{fig:newdatapoint}.

To map the historical prices of both Bitcoin and Litecoin and those of both gold and silver to the same time period, equations~\eqref{eqn:bitcoin_gold_scale}~-~\eqref{eqn:gold_silver_scale} in Fig.~\ref{fig:eqns} are obtained for transforming the absolute times of the prices of either Bitcoin or Litecoin to those of gold and silver. Each linear transformation is determined by the location of the two major peaks in each price history which can be seen in Table~\ref{tab:peaks} and on the price history charts in Figs.~\ref{fig:crypto}~and~\ref{fig:precious}.

{\begin{table}
\centering
\fbox{
\begin{tabular}{p{1.5cm}cccc}
					&\emph{Peak 1}	&\emph{Date}	&\emph{Peak 2}	&\emph{Date}	\\ \hline
\textbf{Bitcoin}	&	213.72		&	09/04/2013	&	1132.26		&	29/11/2013	\\
\textbf{Litecoin}	&	4.84		&	10/04/2013	&	41.689		&	04/12/2013	\\
\textbf{Gold}		&	909.9		&	22/01/1980	&	1901.34		&	05/09/2011	\\
\textbf{Silver}	&	41.5		&	21/01/1980	&	48.46		&	28/04/2011	\\
\end{tabular}}
\caption{\label{tab:peaks}Values of the two most significant peaks in price of Bitcoin, Litecoin, gold and silver.}
\end{table}}

Having scaled the time series to align the major features (peaks), it is necessary to truncate the timescale so as to ensure that the periods of analysis are overlapping. This is done by ensuring that the first data point in the scaled time Bitcoin or Litecoin price history and the first data point in the gold or silver price history have a similar timestamp. Finally, the number of data points in each dataset had to be equal. In order to ensure that this condition was met, the algorithm in Section~\ref{sec:correlation_algorithm}.

\begin{figure}
	\centering
	\makebox[\linewidth][c]{%
	\begin{subfigure}{0.45\textwidth}
		\centering
		\makebox{\includegraphics[width=\textwidth]{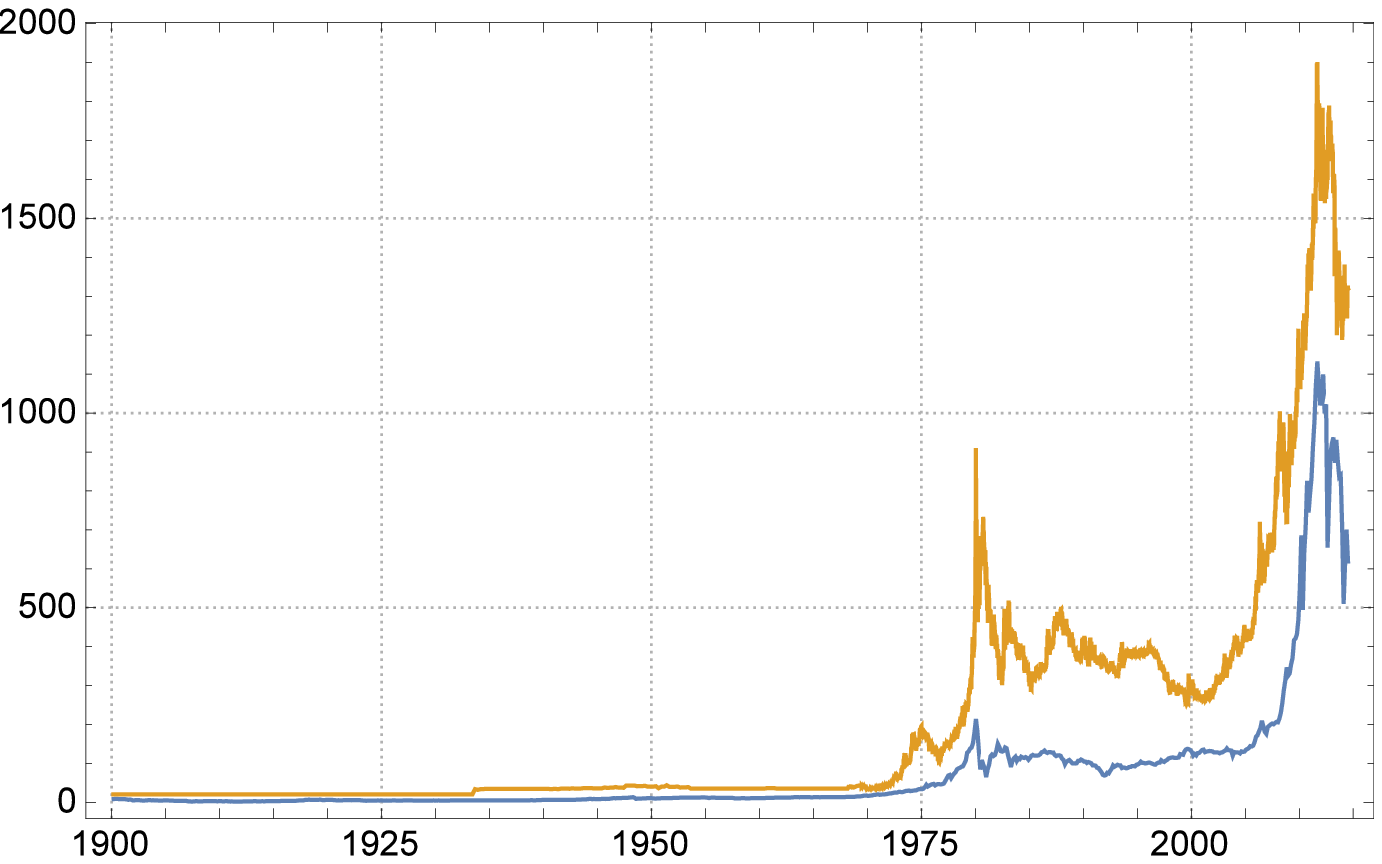}}\medskip \medskip
		\caption{\label{fig:bitcoin_gold_scaled}Historical prices of Bitcoin (blue/dark grey) and gold (orange/light grey) scaled in time.\medskip}
	\end{subfigure}%
	~
	\begin{subfigure}{0.45\textwidth}
		\centering
		\makebox{\includegraphics[width=\textwidth]{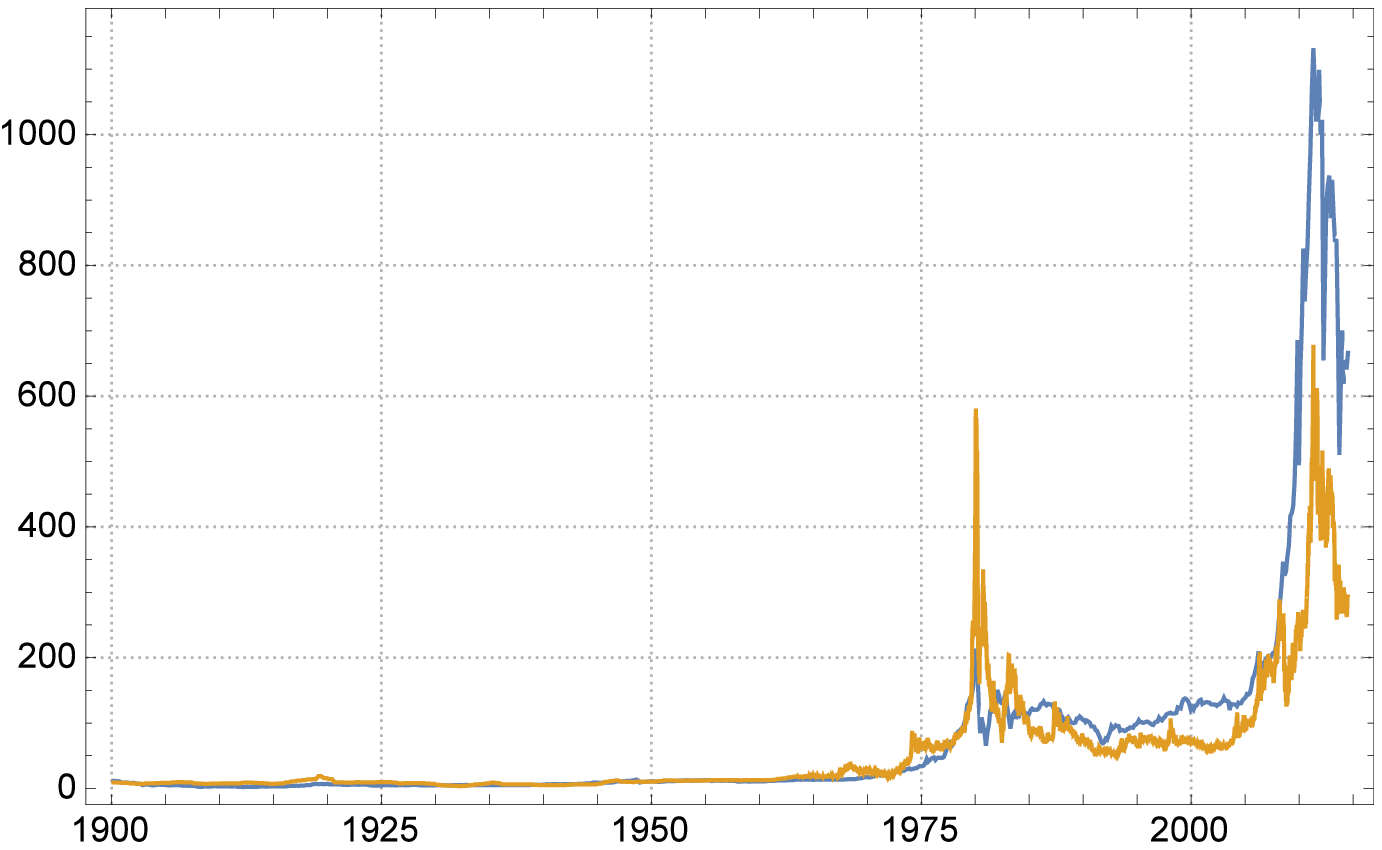}}
		\caption{\label{fig:bitcoin_silver_scaled}Historical prices of Bitcoin (blue/dark grey) and silver (orange/light grey) scaled in time with a $15\times$ price scaling in silver.\medskip}
	\end{subfigure}%
	}\\
	\makebox[\linewidth][c]{%
	\begin{subfigure}{0.45\textwidth}
		\centering
		\makebox{\includegraphics[width=\textwidth]{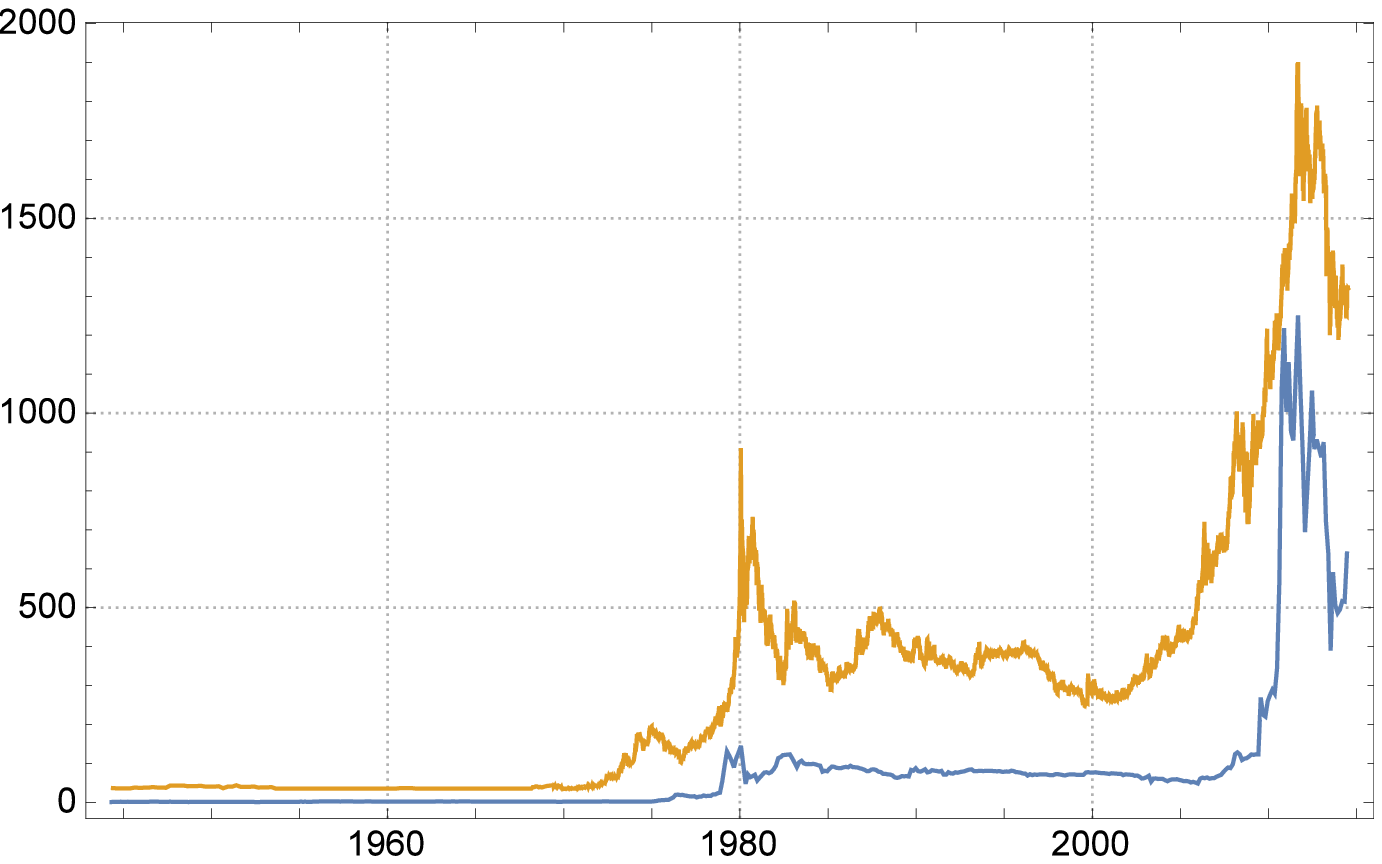}}
		\caption{\label{fig:litecoin_gold_scaled}Historical prices of Litecoin (blue/dark grey) and gold (orange/light grey) scaled in time with a $30\times$ price scaling in Litecoin.\medskip}
	\end{subfigure}%
	~
	\begin{subfigure}{0.435\textwidth}
		\centering
		\makebox{\includegraphics[width=\textwidth]{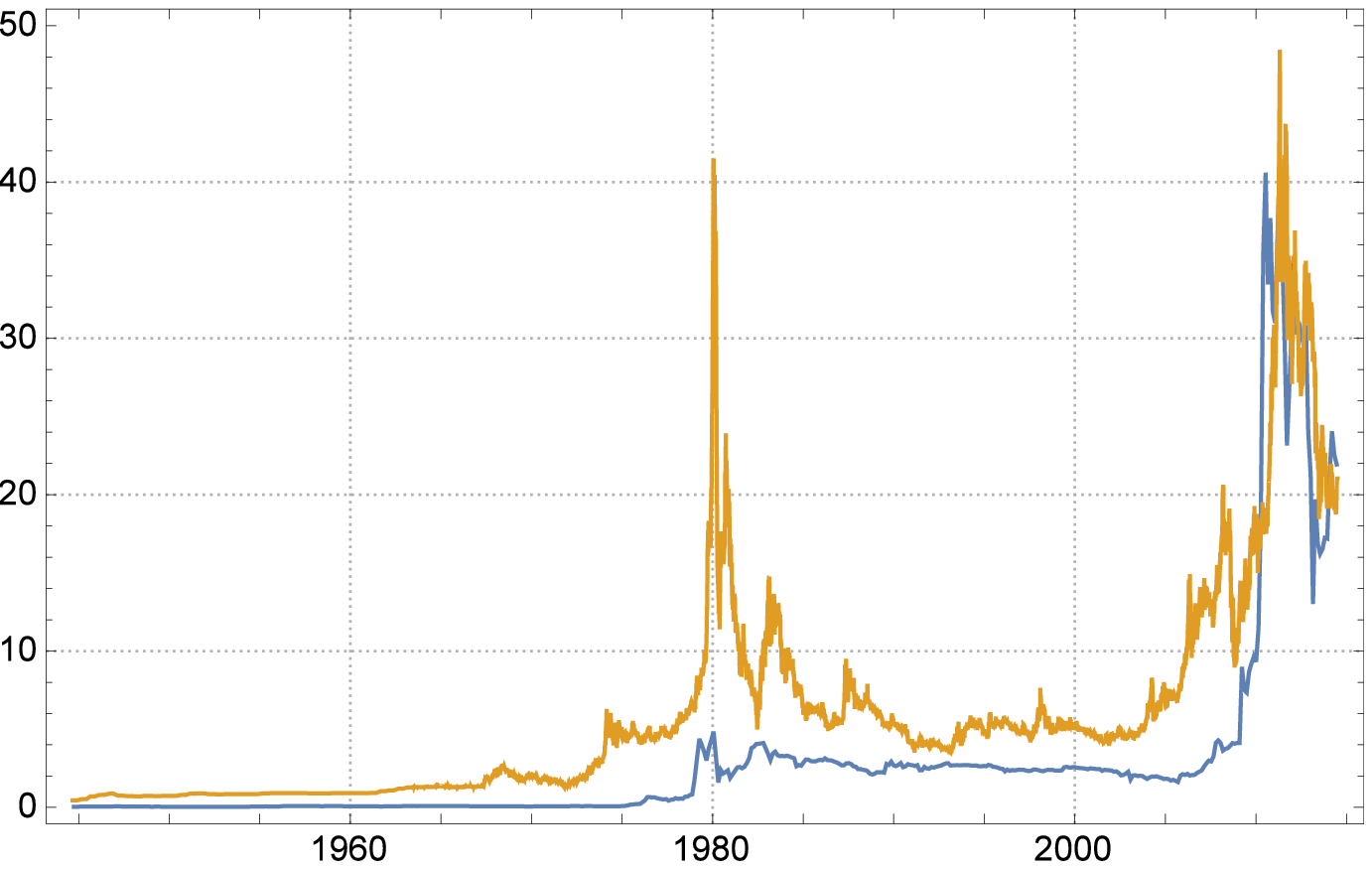}}\medskip
		\caption{\label{fig:litecoin_silver_scaled}Historical prices of Litecoin (blue/dark grey) and silver (orange/light grey) scaled in time.\medskip}
	\end{subfigure}
	}\\
	\makebox[\linewidth][c]{%
	\begin{subfigure}{0.45\textwidth}
		\centering
		\makebox{\includegraphics[width=\textwidth]{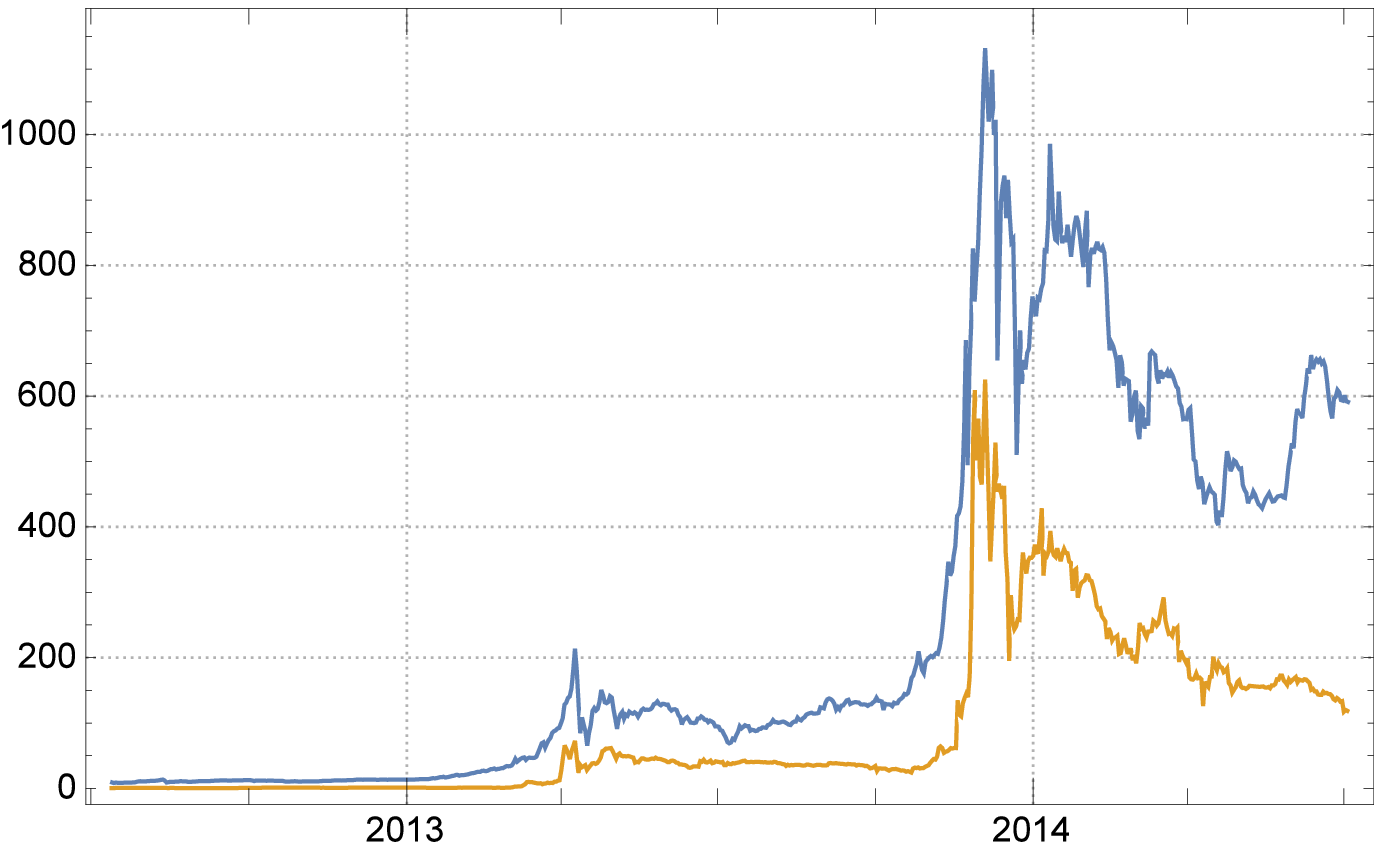}}
		\caption{\label{fig:bitcoin_litecoin_scaled}Historical prices of Bitcoin (blue/dark grey) and Litecoin (orange/light grey) scaled in time with a $15\times$ price scaling in Litecoin.}
	\end{subfigure}%
	~
	\begin{subfigure}{0.45\textwidth}
		\centering
		\makebox{\includegraphics[width=\textwidth]{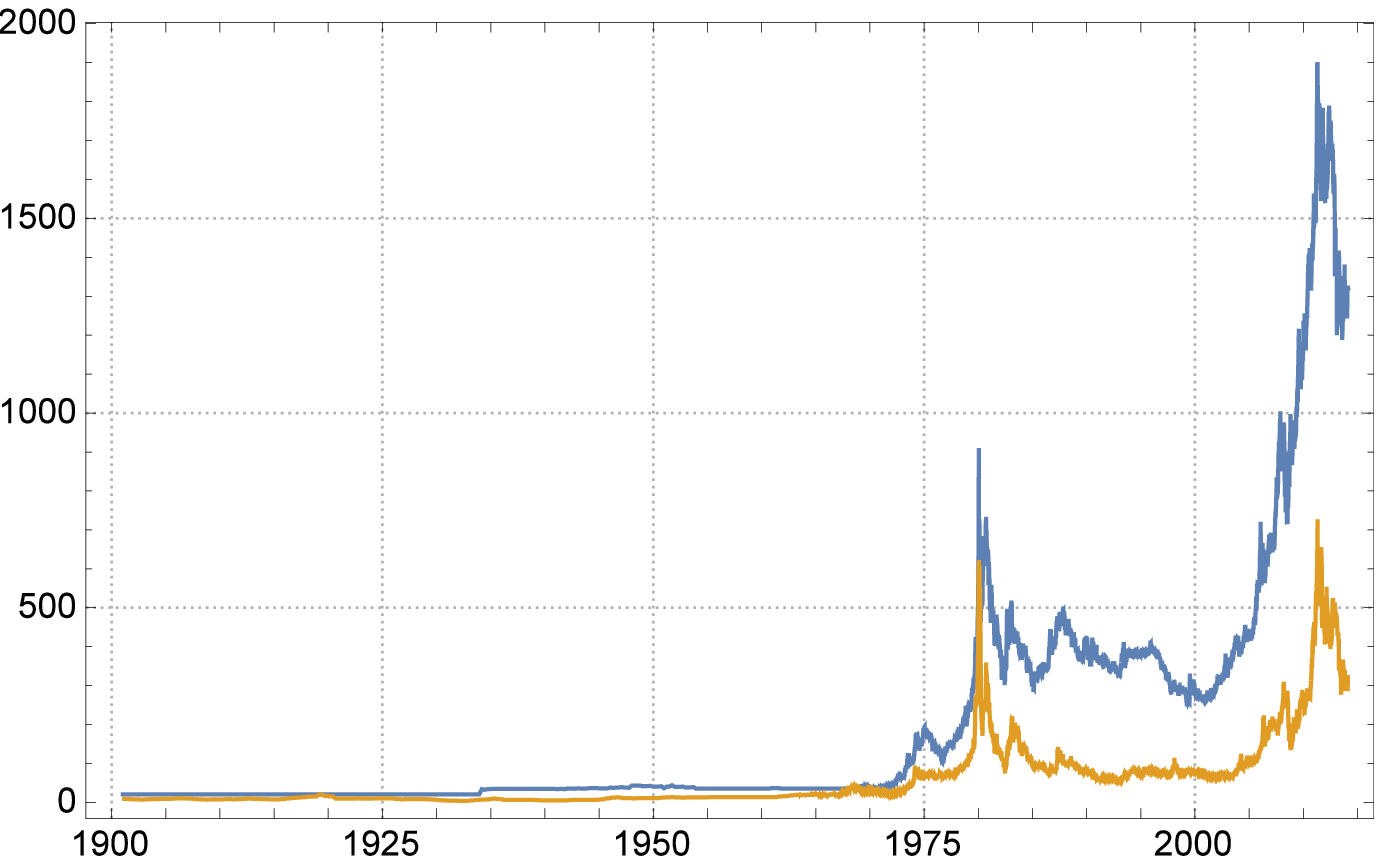}}
		\caption{\label{fig:gold_silver_scaled}Historical prices of gold (blue/dark grey) and silver (orange/light grey) scaled in time with a $15\times$ price scaling in silver.}
	\end{subfigure}
	}
	\caption{\label{fig:scaled}Historical price of cryptocurrencies scaled and concatenated to the same time period as precious metals for visual comparison of similarity in shape.}
\end{figure}

The results when the filtered, scaled and concatenated time series are paired and input into the correlation function in \textit{Mathematica} are depicted in Table~\ref{tab:correlation}.
\begin{figure}
\begin{subequations}
\begin{align}
y	&=	49.3547x-1.7389\times10^{11}		&&	\text{Bitcoin and gold}		\label{eqn:bitcoin_gold_scale}		\\
y	&=	48.8034x-1.71919\times10^{11}	&&	\text{Bitcoin and silver}	\label{eqn:bitcoin_silver_scale}	\\
y	&=	48.3222x-1.70199\times10^{11}	&&	\text{Litecoin and gold}		\label{eqn:litecoin_gold_scale}		\\
y	&=	47.7824x-1.6827\times10^{11}		&&	\text{Litecoin and silver}	\label{eqn:litecoin_silver_scale}	\\
y	&=	49.3547x-1.7389\times10^{11}		&&	\text{Bitcoin and Litecoin}	\label{eqn:bitcoin_litecoin_scale}	\\
y	&=	0.98883x+2.81323\times10^7		&&	\text{Gold and silver}		\label{eqn:gold_silver_scale}
\end{align}
\end{subequations}
\caption{Linear transformations to scale cryptocurrencies' price histories to the same time frame as precious metals'}
\label{fig:eqns}
\end{figure}
After transformation and concatenation of the price histories, the similarity in shape can clearly be seen in Fig.~\ref{fig:scaled} -- prices were scaled in Figs.~\ref{fig:bitcoin_silver_scaled},~\ref{fig:litecoin_gold_scaled},~\ref{fig:bitcoin_litecoin_scaled}~and~\ref{fig:gold_silver_scaled} for visual purposes only.

%{\begin{table}
%\centering
%\fbox{
%\begin{tabular}{l|p{1.2cm}|c|c|c}
%			&	Bitcoin					&	Litecoin	&	Gold	&	Silver	\\ \hline
%Bitcoin		&	\cellcolor{gray!50}	&	\centering 1321 	&	\centering 0.946634		&	\centering 0.906562	\\ \hline
%Litecoin	&	\centering 		&	\cellcolor{gray!50}		&	41.689		&	04/12/2013	\\ \hline
%Gold		&	\centering 0.946634	&	\centering 1231	&	\cellcolor{gray!50}		&	05/09/2011	\\ \hline
%Silver		&	\centering 0.906562	&	21/01/1980	&	48.46		&	\cellcolor{gray!50}		\\
%\end{tabular}}
%\caption{\label{tab:correlation}Values for the correlation between movements in the price of pairs of Bitcoin, Litecoin, Gold and Silver}
%\end{table}}

{\begin{table}
\centering
\fbox{
\begin{tabular}{p{1.5cm}|p{1.5cm}p{1.5cm}p{1.5cm}p{1.5cm}}
					&	\textbf{Silver	}		&	\textbf{Gold}			&	\textbf{Litecoin}		&	\textbf{Bitcoin}		\\ \hline
\textbf{Bitcoin}	&	0.906562				&	0.946634				&	0.976853				&	\cellcolor{gray!50}	\\
\textbf{Litecoin}	&	0.841576				&	0.872573				&	\cellcolor{gray!50}	&							\\
\textbf{Gold}		&	0.911728				&	\cellcolor{gray!50}	&							&							\\
\textbf{Silver}	&	\cellcolor{gray!50}	&							&							&							\\
\end{tabular}}
\caption{\label{tab:correlation}Values for the correlation between movements in the price of pairs of Bitcoin, Litecoin, gold and silver.}
\end{table}}

\begin{figure}
	\centering
	\makebox[9cm][l]{%
	\begin{subfigure}{0.4\textwidth}
		\centering
		\makebox{\includegraphics[width=\textwidth]{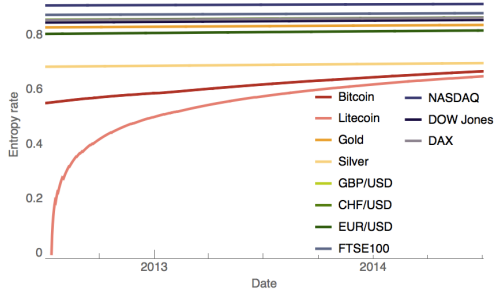}}
		\caption{\label{fig:marketcompressibility} Normalised block entropy of discretised prices of markets time series for the last two years from July 2012 to July 2014.}
	\end{subfigure}%
	}
	\makebox[4.5cm][r]{%
	\begin{subfigure}{0.4\textwidth}
		\centering
		\makebox{\includegraphics[width=\textwidth]{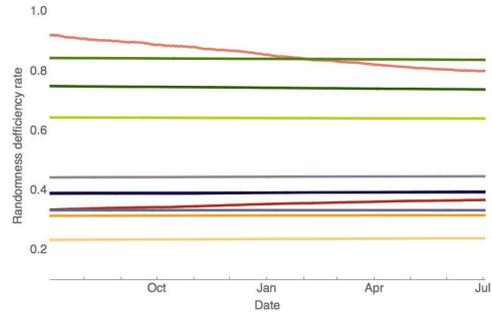}}
		\caption{\label{fig:marketentropy} Normalised compressibility of real-value prices of markets time series from July 2013 to July 2014.\bigskip}
	\end{subfigure}%
	}\\ \bigskip \bigskip
	\centering
	\makebox[9cm][l]{%
	\begin{subfigure}{0.4\textwidth}
		\centering
		\makebox{\includegraphics[width=\textwidth]{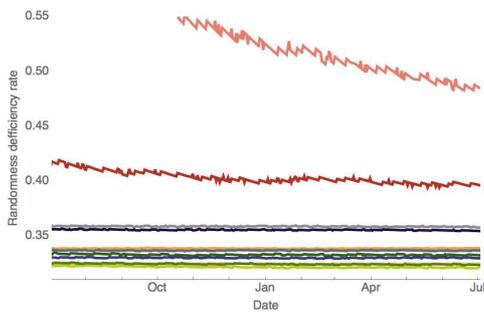}}
		\caption{\label{fig:marketcompressibility} Compressibility of discretised prices of markets time series from July 2013 to July 2014.}
	\end{subfigure}%
	}
	\makebox[4.5cm][r]{%
	\begin{subfigure}{0.4\textwidth}
		\centering
		\makebox{\includegraphics[width=\textwidth]{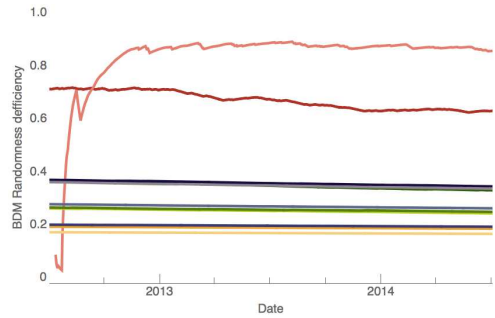}}
		\caption{\label{fig:marketbdmdeff} Normalised BDM randomness deficiency of discretised prices of markets time series for the last two years from July 2012 to July 2014.}
	\end{subfigure}%
	}\\ \bigskip \bigskip

	\centering
	\makebox[\linewidth][c]{%
	\begin{subfigure}{0.4\textwidth}
		\centering
		\makebox{\includegraphics[width=\textwidth]{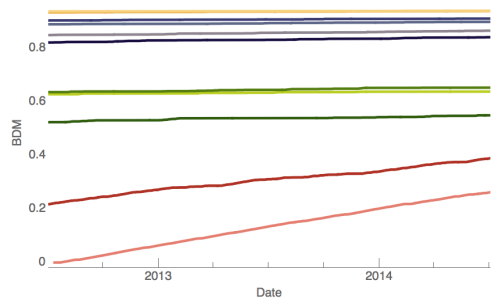}}
		\caption{\label{fig:marketbdm} Normalised BDM of discretised prices of markets from July 2012 to July 2014.}
	\end{subfigure}%
	}\\	
	\caption{\label{fig:compressandentropy} Compressibility and block entropy of three mature markets (stock, exchange and two precious metals) and the two main cryptocurrencies markets. (colours in original online version)}
\end{figure}

\subsection{Entropy and Compressibility}

\begin{figure}
	\centering
	\makebox{\includegraphics[width=0.75\textwidth]{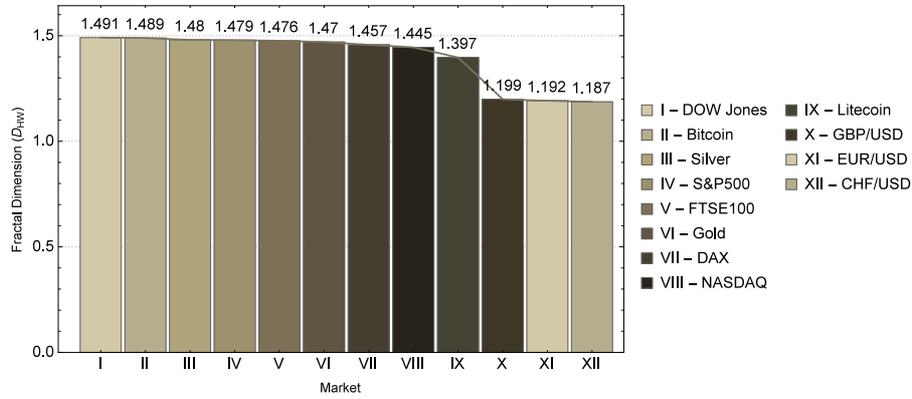}}
	\caption{\label{fig:fractal_dimension}Fractal dimension calculated using Hall-Wood estimator of all investment products in Table~\ref{tab:comps} using entire price histories.}
\end{figure}

\begin{figure}
	\centering
	\makebox{\includegraphics[width=0.75\textwidth]{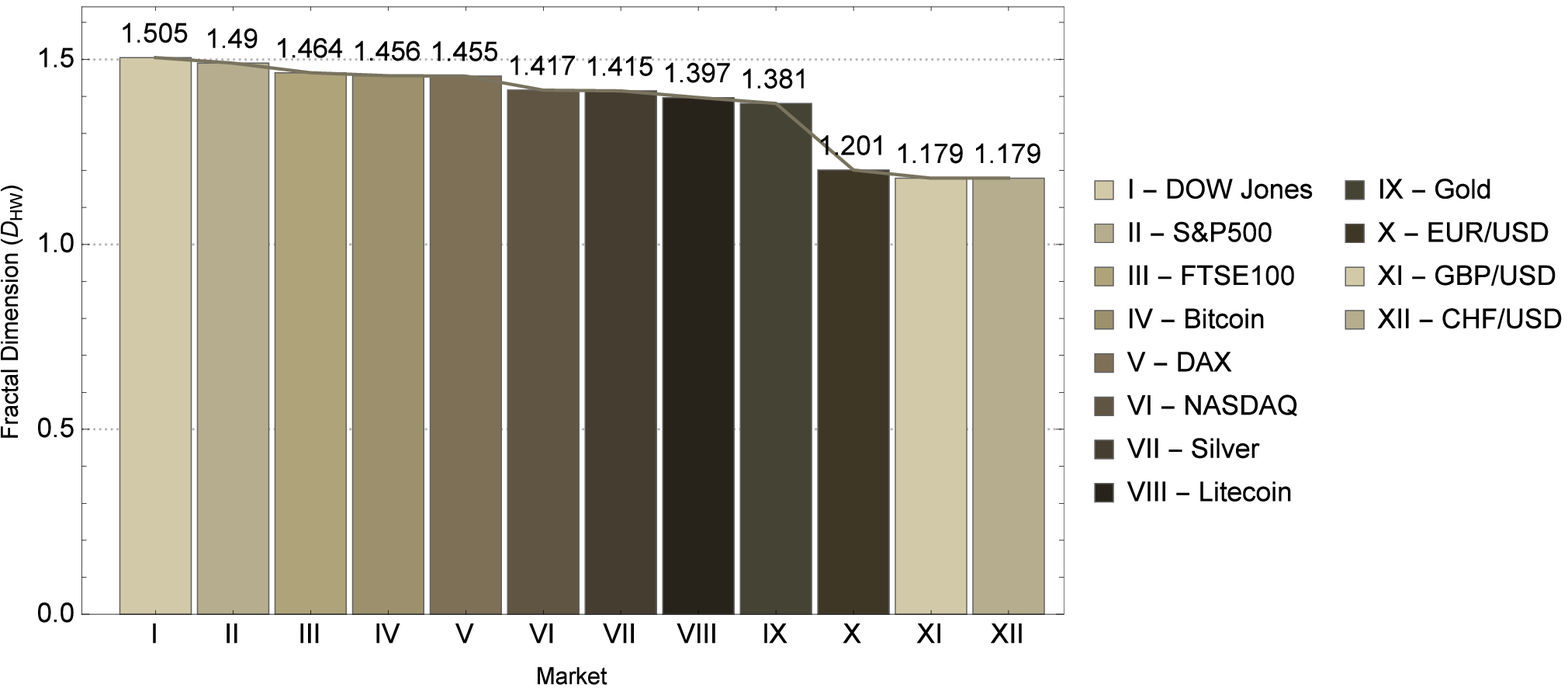}}
	\caption{\label{fig:fractal_dimension2}Fractal dimension calculated using Hall-Wood estimator of all investment products in Table~\ref{tab:comps} using price histories since 13/07/2012.}
\end{figure}

When analysing compressibility, in Fig.~\ref{fig:compressandentropy} it can be seen that the cryptocurrencies are placed in two very different locations; Bitcoin is often positioned close or among the stock indices whilst Litecoin, with relatively much lower volume, is currently placed alongside the foreign exchange markets. There are, however indications that Litecoin may be moving towards Bitcoin's trend closer to the precious metals markets and stock indices. There is also a stability of the mature markets and similar behaviour of their indices. A variation metric defined as \emph{randomness deficiency} is simply the compression or entropy value divided by the length of the subtime series. This gives a ratio of how removed the value of the sequence in question is from its maximum complexity represented by the full uncompressed length of the same sequence.

On the one hand, the block entropy consists in taking the sum of the entropies of individual initial segments of the price time series. More formally,
$$
H\left(X\right) = \sum_{i=1}^{n} H\left(X_i\right),
$$
where $n$ can run up to the length of the time series, or a small number, in this case 4, as it is computationally expensive, but can see longer correlations than entropy on digits alone ($i = 1$). All entropy results were calculated by block entropy unless otherwise stated. On the other hand, the \emph{Block Decomposition Method} (BDM), by nature, only works on discretised values whilst in Fig.~\ref{fig:compressandentropy}b compression was applied directly to  real-value prices. In all other cases the binarised time series encoding ups and downs of price motion was used.

\subsection{Fractal Dimension}

Using the Hall-Wood estimator, the fractal dimension of the price histories of all the investment products listed in Table~\ref{tab:comps} can be seen visualised in Fig.~\ref{fig:fractal_dimension}. Similar to the compressibility and entropic tests, fractal dimension displays two clear groups separating the currency exchange market from the stock and precious metals, but converges in placing Bitcoin and Litecoin in different places, closer to currencies for the latter and among the stock market for the former, hence in full agreement to compressibility and close agreement with block entropy.

When the fractal dimension is calculated for the period since 13/07/2012, the date at which the Litecoin price history starts, the grouping is again very clear and places Bitcoin among stock indices and Litecoin closer to foreign exchange, as can be seen in Fig.~\ref{fig:fractal_dimension2}.

A completely non-random graph (i.e. a straight line) has fractal dimension of 1 and the dimension, $d$ of a line can take values $d\in\left[1,2\right)$. The closer the fractal dimension to 1, the less random the line, a symmetric random walk with holding probability 0 has equal probability of either rising or falling, and as such has fractal dimension 1.5. This can be interpreted as meaning that a time series with fractal dimension $d\in\left[1,1.5\right)$ is less random (i.e. less rough) than a random series and if $d\in\left(1.5,2\right)$ then it is more random (i.e. more rough) than a random series.

When the entire price history is used to calculate the fractal dimension, every product's price history fractal dimension is between 1 and 1.5. The majority of the products (cryptocurrencies, precious metals and stock indices) - group 1 - have dimensions ranging between 1.397 and 1.491, whilst the foreign exchange - group 2 - have fractal dimensions ranging from 1.187 to 1.199. In comparison, when the price histories from 13/07/2012 are used, group 1's fractal dimensions range from 1.381 to 1.505 whilst group 2's fractal dimensions range from 1.179 to 1.201. This shows that regardless of whether the whole price history is used or not, foreign exchange prices are significantly less rough than those of cryptocurrencies, precious metals and stock indices. 

\section{Conclusions}

Using statistical, information-theoretic, algorithmic and fractal measures we have analysed the behaviour of various markets in connection to cryptocurrencies. We have shown that Bitcoin has some similarities to precious metal markets -- particularly gold and silver -- when looking at how removed their price movement distributions are from lognormal. Using algorithmic and fractal measures, we have also shown that Bitcoin often displays similar behaviour to stock and precious metal markets appearing to be akin to a hybrid instrument - on the surface a currency but with property transaction behaviour. Most classification results from the measures of complexity used such, as in ~\ref{fig:compressandentropy}, assign mature markets and the three exchange rates (green colours) the lowest (normalised) compressibility, followed by stock (various colours) and metals (gold). It is clear that Litecoin is moving fast whilst volume increases in Fig.~\ref{fig:compressandentropy} and according to most complexity measures cryptocurrencies come closer to stock and the precious metal markets.

We have shown that there is a clear distinction between the complexity and fractal roughness between the three families of markets studied herein; stock, foreign exchange and precious metals. We found that Bitcoin displays one of the most complex and roughest behaviours across all markets, whilst Litecoin is sometimes closer to currencies but moving in the direction of Bitcoin. This is most likely due to its low volume, as it has a clear trend towards the position of Bitcoin over a period of fast transaction volume increase. Indeed, smaller cryptocurrencies seem too weak to carry signals to show the behaviour displayed by Bitcoin, even though they clearly follow Bitcoin's trends. To the authors knowledge this is the first time that complexity measures of different nature -- namely information theoretic, algorithmic complexity, fractal and statistical -- converge at clustering the behaviour of complex systems such as markets, and have been applied to quantifying common similarities and dissimilarities among them. 

\section*{Acknowledgments}

We want to thank staff from the Wolfram Science Summer School, particularly to Stephen Wolfram, Todd Rowland and Jason Cawley. % Additionally, Anastasia Papavasilliou for her input on Signatures of Rough paths.

\bibliography{mybib}
\bibliographystyle{chicago}

\end{document}